\documentclass{ws-ijmpe}
\usepackage[super,compress]{cite}
\begin{document}

\markboth{Guang-You Qin}{Anisotropic Flow and Jet Quenching in Relativistic Nuclear Collisions}

%%%%%%%%%%%%%%%%%%%%% Publisher's Area please ignore %%%%%%%%%%%%%%%
\catchline{}{}{}{}{}
%%%%%%%%%%%%%%%%%%%%%%%%%%%%%%%%%%%%%%%%%%%%%%%%%%%%%%%%%%%%%%%%%%%%

\title{Anisotropic Flow and Jet Quenching in Relativistic Nuclear Collisions}

\author{\footnotesize Guang-You Qin}

\address{Institute of Particle Physics and Key Laboratory of Quark and Lepton Physics (MOE), \\ Central China Normal University, Wuhan, 430079, China
 \\
guangyou.qin@mail.ccnu.edu.cn}

\maketitle

\begin{history}
\received{Day Month Year}
\revised{Day Month Year}
%\accepted{Day Month Year}
%\comby{(xxxxxxxxxx)}
\end{history}

\begin{abstract}

The exploration of the strong-interaction matter under extreme conditions is one of the main goals of relativistic heavy-ion collisions. We provide some of the main results on the novel properties of quark-gluon plasma, with particular focus given to the strong collectivity and the color opaqueness exhibited by such hot and dense matter produced in high-energy nuclear collisions at RHIC and the LHC.

\end{abstract}


\ccode{PACS numbers:}

%\tableofcontents

\section{Introduction}

The study of QCD matter under extreme conditions of temperatures and densities is of great interest in the field of high-energy nuclear physics.
The simulations from lattice QCD have demonstrated that as the temperature of the QCD matter is increased above a certain value $T_c  \sim 155$~MeV \cite{Borsanyi:2010bp, Bazavov:2011nk, Bhattacharya:2014ara}, quarks and gluons that are confined in normal hadronic matter will be liberated, and form a novel state of matter which is usually referred to as quark-gluon plasma (QGP).
QGP is believed to have existed at a few micro-microseconds after the Big Bang, when the temperature of the Universe was so high that quarks and gluons could not bind together.
Such hot and dense nuclear matter may be created in the laboratories by colliding two heavy nuclei at ultra-relativistic energies, such as those performed at the Relativistic Heavy-Ion Collider (RHIC) at Bookhaven National Laboratory (BNL) and the Large Hadron Collider (LHC) at European Organization for Nuclear Research (CERN).
The main goal of high-energy heavy-ion experiments is to explore various novel properties of the hot and dense QGP, and map out the phase structure of the strong-interaction matter.
With the center-of-mass collision energies up to 200~GeV per nucleon pair at RHIC and more than one order of magnitude higher at the LHC, the highest temperatures of the hot matter created in these experiments can reach $\sim 370-470$~MeV  \cite{Heinz:2013th, Gale:2013da, Song:2013gia, Romatschke:2009im, Huovinen:2013wma}, which is well above the transition temperature predicted by the lattice QCD calculations \cite{Borsanyi:2010bp, Bazavov:2011nk, Bhattacharya:2014ara}.

\begin{figure}
\begin{center}
\includegraphics*[width=12.0cm]{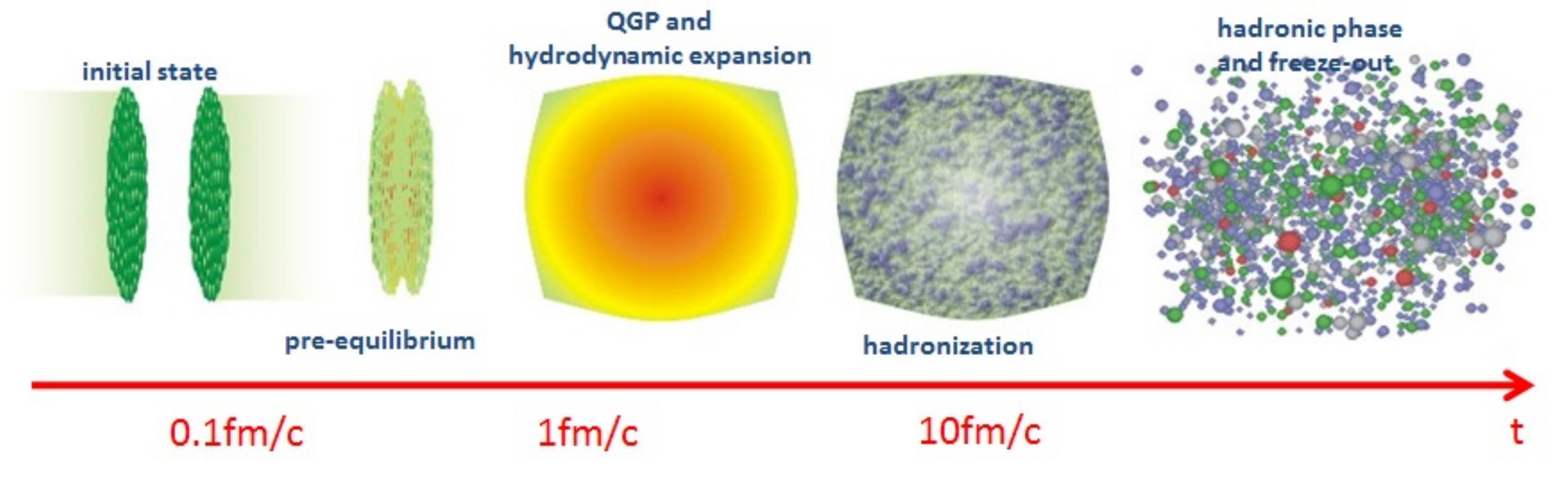}
\caption{(Color online) Illustration of different dynamical evolution stages in a typical heavy-ion collision.
}
\label{fig_bass_hic}
\end{center}
\end{figure}

Our current understanding of a typical heavy ion collision is schematically illustrated in Fig. \ref{fig_bass_hic}.
One starts from two fast-moving nuclei heading on each other.
In the laboratory frame, the two colliding nuclei look like pancakes due to the Lorentz contraction.
After two nuclei penetrate through each other, a large amount of energy is deposited in the central collision zone and a highly-excited nuclear matter is produced.
The produced matter experiences a pre-equilibrium evolution stage, and then reaches (a certain degree of) local thermal equilibrium.
The equilibrated hot and dense QGP matter then expands like a relativistic fluid and cools down.
When the temperature/density of the fluid drops below a certain value, it undergoes a transition from QGP to hadronic phase.
The produced hadrons continue to interact with each other until the kinetic freezout, and then fly to the detectors.
One can see that the deconfined color degrees of freedom in the QGP matter produced in energetic heavy-ion collisions are not directly observable, but manifest in the final state hadrons and other colorless objects.
The essential task in the study of hot and dense QGP using relativistic heavy-ion collisions is to find clear and unambiguous connections between the produced QGP and the final (hadronic) observables, and to identify reliable signatures for the formation of QGP.

Experimental results have shown that the QGP matter produced in these energetic nuclear collisions exhibits many remarkable properties that differ from anything that has been seen before \cite{Gyulassy:2004zy,Adcox:2004mh,Arsene:2004fa,Back:2004je,Adams:2005dq,Muller:2006ee,Muller:2012zq}.
On one hand, such hot and dense nuclear matter shows strong (anisotropic) collective behavior.
The successfulness of relativistic viscous hydrodynamics in the description of the space-time evolution of the bulk matter, especially the observed anisotropic collective flow, indicates that the produced QGP matter is strongly interacting and has very small specific shear viscosity.
On the other hand, the produced matter shows strong opaqueness to the attenuation of high energy partons; this is usually called jet quenching.
Model calculations based on parton energy loss which originates from the interaction between the propagating hard parton and the hot and dense QCD medium can explain quite well many phenomena associated with hard jets.
These and other exciting findings from RHIC and the LHC indicate that the studies of strong-interaction matter in high energy nuclear collisions have entered a completely new era.

In this report, we mainly focus on two most important phenomena as mentioned above, i.e., anisotropic collective flow and jet quenching in relativistic nuclear collisions.
In particular, we will discuss the recent efforts on the quantitative determination of two important transport coefficients, namely, the shear viscosity to entropy ratio (the specific shear viscosity) $\eta/s$ and the scaled jet quenching parameter $\hat{q}/T^3$ (or $\hat{q}/s$), via systematic phenomenological studies of soft particles produced from the bulk matter and high transverse momentum particles originating from hard partonic jets at RHIC and the LHC.
The precise determination of these two transport coefficients and their temperature dependences can provide us a quantitative estimation of the interaction nature of the hot and dense nuclear matter produced in heavy-ion collisions, e.g., how a weakly-coupled quark-gluon matter at sufficiently high temperatures turns into a strongly-coupled relativistic fluid at the energies achieved at RHIC and the LHC experiments \cite{Majumder:2007zh}.

\section{Collectivity and Fluidity of QGP}

\begin{figure}
\begin{center}
\includegraphics*[width=12.0cm]{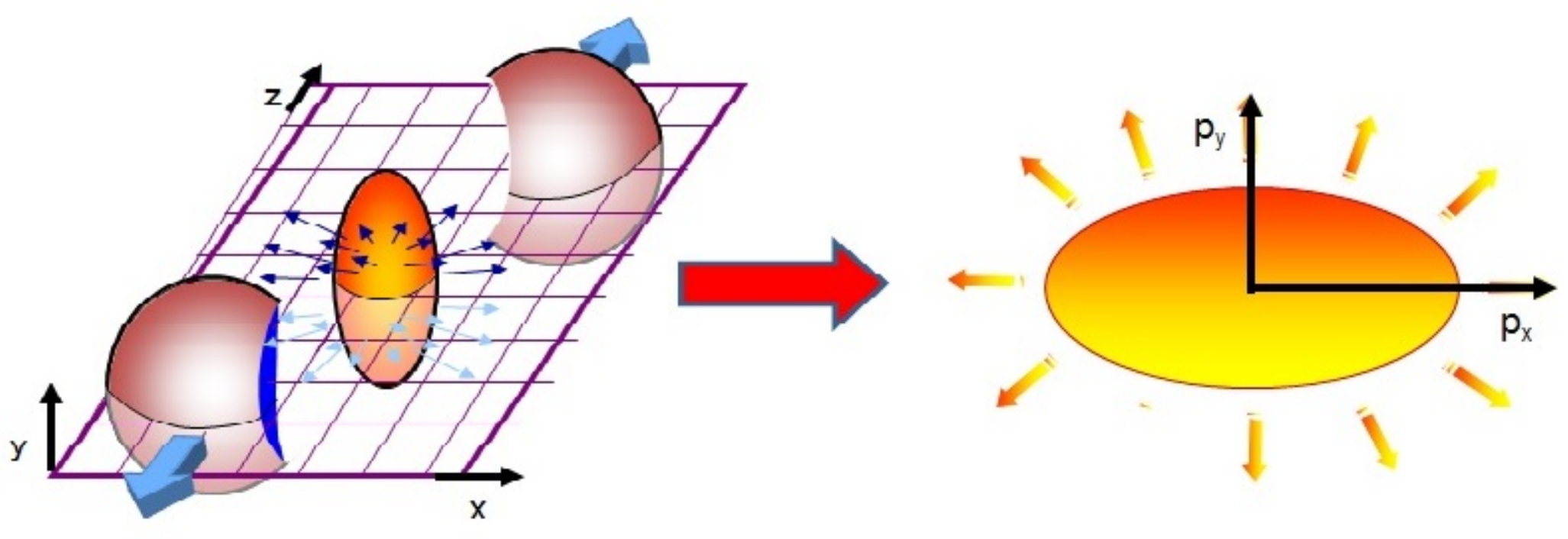}
\caption{(Color online) Illustration a typical non-central heavy-ion collision in which the eccentric shape of initial collision geometry is translated into the elliptic flow anisotropy in the final state momentum space.
}
\label{fig_e2_v2}
\end{center}
\end{figure}

In a typical non-central nucleus-nucleus (A-A) collision, the produced hot and dense nuclear matter is anisotropic in the plane transverse to the bean direction (see Fig. \ref{fig_e2_v2}).
This geometric anisotropy is usually quantified by the second-order eccentricity $\epsilon_2 = \langle y^2 - x^2\rangle /\langle y^2 + x^2\rangle $.
Due to the interaction among the medium constituents, the initial geometric anisotropy will be converted into the anisotropy of the final state particle momentum distribution, which is usually quantified by the elliptic flow coefficient $v_2 = \langle p_x^2 - p_y^2\rangle /\langle p_x^2 + p_y^2\rangle $ \cite{Ollitrault:1992bk,Adler:2003kt, Adams:2003am, Aamodt:2010pa, ATLAS:2011ah, Chatrchyan:2012ta}.
The observed large elliptic anisotropy and the dynamical evolution of the hot and dense QGP matter have been well described by the simulations from relativistic hydrodynamics \cite{Rischke:1995ir, Rischke:1995mt, Kolb:2000fha, Teaney:2000cw, Huovinen:2001cy, Hirano:2002ds, Nonaka:2006yn, Romatschke:2007mq, Dusling:2007gi, Molnar:2008xj, Song:2007ux, Niemi:2008ta, Luzum:2008cw, Schenke:2010rr, Shen:2014vra}.
This has led to the conclusion that at the energies achieved at RHIC and the LHC the produced QGP is a strong-coupled nuclear matter, which behaves like a relativistic fluid.

\subsection{Relativistic hydrodynamics and QGP viscosity}

In relativistic hydrodynamics (with no net conserved charge), one solves the conservation equation for the energy-momentum tensor,
\begin{eqnarray}
\partial_\mu T^{\mu\nu} = 0 \,,
\end{eqnarray}
where $T^{\mu\nu}(x)$ is given by
\begin{eqnarray}
T^{\mu\nu}(x) = e(x)u^\mu(x) u^\nu(x) - P(x) \Delta^{\mu\nu}(x) - \Pi(x) \Delta^{\mu\nu}(x) + \pi^{\mu\nu}(x) \,.
\end{eqnarray}
Here $e(x)$ and $P(x)$ are the local energy density and pressure of the fluid.
$u(x)=\gamma[1,\mathbf{v}(x)]$ is the local four-velocity ($u^\mu u_\mu = 1$), and $\Delta^{\mu\nu}(x) = g^{\mu\nu} - u^{\mu}(x) u^{\nu}(x)$.
$\Pi(x)$ and $\pi^{\mu\nu}(x)$ are the bulk pressure and the shear tensor.
The first two terms are for ideal hydrodynamics, and the last two terms are viscous corrections.

One can see that there are $5$ independent variables in ideal hydrodynamics ($e$, $P$ and $\mathbf{v}$), and $6$ more unknowns in the viscous correction terms ($\Pi$ and $5$ independent components for $\pi^{\mu\nu}$).
The evolution equation for the energy-momentum tensor renders 4 equations (for the energy density $e$ and the fluid velocity $\mathbf{v}$).
The equation of state (EoS) will relate the pressure to the energy density, $P=P(e)$, which is usually taken from lattice QCD calculation.
For viscous hydrodynamics, additional evolution equations for the bulk pressure and the shear tensor are needed.
They are usually obtained using the entropy principle, or derived from the kinetic theory via the moment method or the gradient expansion \cite{Israel:1976tn, Israel:1979wp, Baier:2007ix, Romatschke:2009kr, El:2009vj, Denicol:2012cn, Jaiswal:2013fc, Bazow:2013ifa}.
In the Israel-Stewart theory \cite{Israel:1976tn, Israel:1979wp}, we have:
\begin{eqnarray}
&& D \Pi = - \frac{1}{\tau_\Pi} \left[\Pi + \zeta \partial_\alpha u^{\alpha} +\Pi  \zeta T  \partial_\alpha\left(\frac{\tau_\Pi}{2\zeta T} u^\alpha\right) \right] \,,\nonumber\\
&& \Delta^{\mu}_{\alpha} \Delta^{\nu}_{\beta} D \pi^{\alpha\beta} = - \frac{1}{\tau_\pi} \left[\pi^{\mu\nu} - 2\eta \sigma^{\mu\nu} +  \pi^{\mu\nu}  \eta T \partial_\alpha\left(\frac{\tau_\pi}{2\eta T} u^\alpha\right) \right] \,.
\end{eqnarray}
Here $D = u^\mu \partial_\mu$ is the comoving derivative, $\sigma^{\mu\nu} = \frac{1}{2}(\nabla^\mu u^\nu + \nabla^\nu u^\mu) - \frac{1}{3} \Delta^{\mu\nu} \nabla_\alpha u^\alpha$,
and $\nabla^{\alpha} = \Delta^{\alpha\beta}\partial_{\beta}$ is the local spatial derivative.
$\zeta$, $\eta$ are bulk viscosity and shear viscosity, and $\tau_\Pi$, $\tau_\pi$ are the corresponding relaxation times.
Currently most hydrodynamics calculations ignore the bulk viscosity effects since shear viscosity provides the dominant contribution regarding the influence of the flow development in ultra-relativistic heavy-ion collisions \cite{Song:2009rh}.

In order to solve the relativistic hydrodynamics evolution equations,  we need the spatial distribution of the energy momentum tensor at the starting time $\tau_0$ of the hydrodynamic evolution, which is usually obtained from various initial condition models.
After solving hydrodynamic equations, one may obtain the spectra of the particles produced from the bulk matter according to the Coope-Frye description \cite{Cooper:1974mv}:
\begin{eqnarray}
E\frac{dN_i}{d^3p} = \frac{dN_i}{d^2p_Tdy} = \frac{g_i}{(2\pi)^3} \int_{\Sigma} p^{\mu}d\Sigma_\mu(x) f_i(x,p) \,,
\end{eqnarray}
where $d\Sigma_\mu(x)$ is an infinitesimal element whose direction is normal to the hypersurface $\Sigma(x)$, $g_i$ is the degeneracy factor, and $f_i(x,p)$ is the phase space distribution of particle species $i$.

\begin{figure}
\begin{center}
\includegraphics*[width=12.0cm]{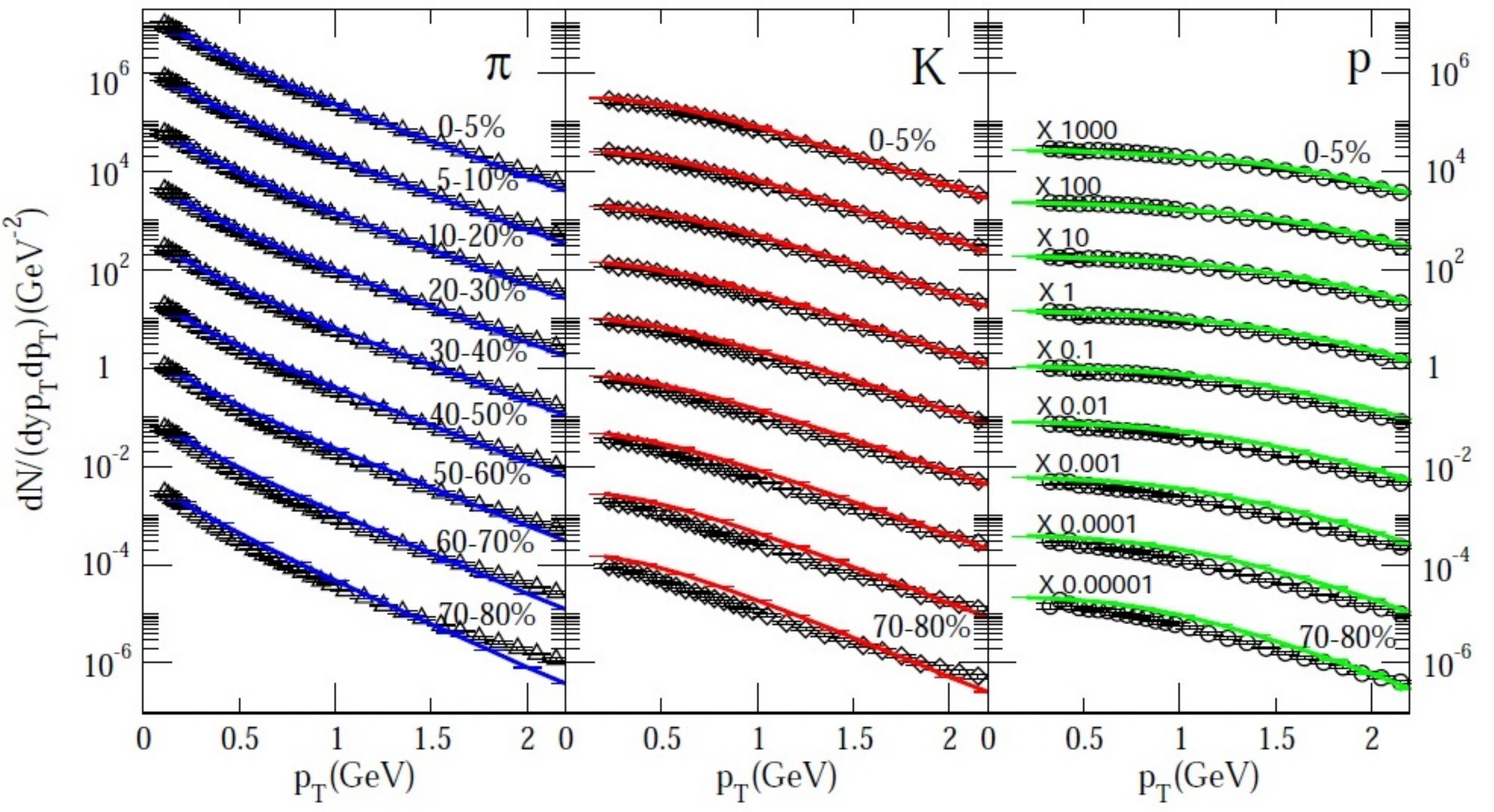}
\caption{(Color online) %Upper panels:
The $p_T$ spectra for pions, kaons and protons in Pb-Pb collisions at $2.76$~ATeV at the LHC.
From top to bottom the curves are for 0-5\%, 5-10\%, 10-20\%, 20-30\%, 30-40\%, 40-50\%, 50-60\%, 60-70\%, 70-80\% collision centralities (for clearer presentation, the spectra are multiplied by a factor of $1000$, $100$, $10$, $1$, $0.1$, $0.01$, $0.001$, $10^{-4}$ and $10^{-5}$, respectively).
Theoretical curves are from VISHNU \cite{Song:2013qma} and experimental data are from ALICE \cite{Abelev:2013vea}.
The figures are taken from Ref. \cite{Song:2013qma}.
}
\label{fig_song_spectra}
\end{center}
\end{figure}

Fig. \ref{fig_song_spectra} shows the $p_T$ spectra for identified hadrons (pions, kaons and protons) in Pb-Pb collisions at $2.76$~ATeV at the LHC.
Theoretical curves are from VISHNU calculations \cite{Song:2013qma} and experimental data are from ALICE measurements \cite{Abelev:2013vea}.
Here the QGP specific shear viscosity $\eta/s$ in VISHNU calculations is taken to be a constant $\eta/s=0.16$ and the corresponding relaxation time is set as $\tau_\pi = 3\eta/(sT)$.
The equation of stateis taken from s95p-PCE parameterization which has been constructed by matching the lattice QCD calculation at higher temperature to a chemically frozen hadron resonance gas at lower temperature \cite{Huovinen:2009yb}.
The initial time for hydrodynamic evolution is set to be $\tau_0 = 0.9$~fm/c, and the initial entropy density profiles are generated from the MC-KLN model \cite{Drescher:2006ca}.
One can see that except for the most peripheral collisions, where one does not expect hydrodynamics model to work well, the VISHNU model can provide a good description of the ALICE data for all three particle species over a variety of the collision centralities.

Much of recent attention in the hydrodynamic studies of heavy-ion collisions has been paid to the quantitative extraction of transport coefficients such as the shear viscosity to entropy ratio $\eta/s$ of the produced hot and dense QGP matter.
Shear viscosity reflects the ability of the matter to transport momentum between different parts of the system.
From kinetic theory, it is related to the mean free path $\lambda$ or the transport cross section $\sigma_{\rm tr}$ as: $\eta = \frac{1}{3}\rho\langle p \rangle \lambda = \frac{1}{3} \langle p \rangle / \sigma_{\rm tr}$, where $\langle p \rangle$ is the average momentum carried by the matter constituents.
Thus shear viscosity provides a direct measure of the interaction strength among the matter constituents.

\begin{figure}
\begin{center}
\includegraphics*[width=12.0cm]{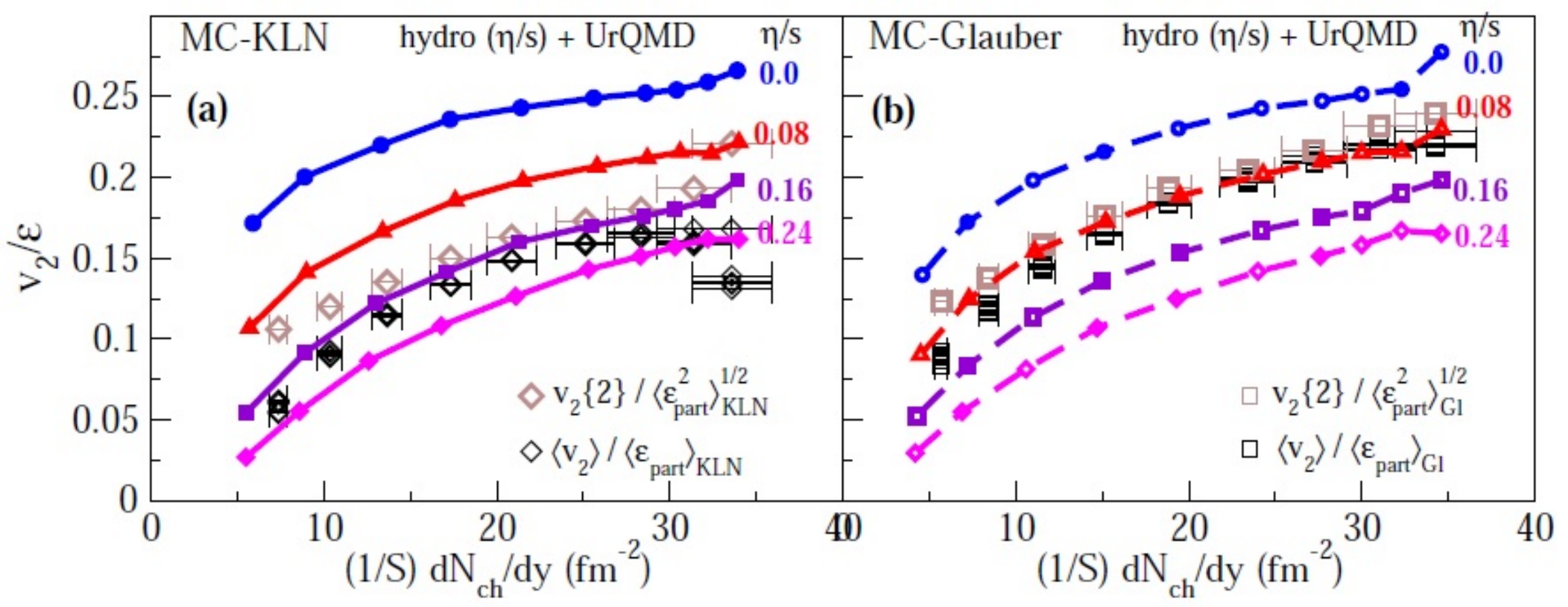}
\caption{(Color online) The integrated elliptic flow $v_2$ for all charged hadrons scaled by the eccentricity $\epsilon_2$ in Au-Au collisions at $200$~AGeV as a function of the charged hadron multiplicity per unit overlap transverse area (from Ref. \cite{Song:2010mg}).
}
\label{fig_song_v2_viscosity}
\end{center}
\end{figure}

Fig. \ref{fig_song_v2_viscosity} shows an effort to extract the specific shear viscosity $\eta/s$ from the charged hadron elliptic flow $v_2$ data in Au-Au collisions at RHIC \cite{Song:2010mg}.
The figure shows the $p_T$ integrated elliptic flow $v_2$ scaled by the initial eccentricity $\epsilon_2$ as a function of collision centrality [represented by the charged hadron multiplicity density per unit overlap area $(1/S)(dN_{\rm ch}/dy)$].
The theoretical curves are from the VISHNU calculations \cite{Song:2010mg} using two different initial condition models: MC-Glauber \cite{Miller:2003kd} (left panel) and MC-KLN \cite{Drescher:2006ca} (right panel).
The experimental data show the elliptic flow measured using two different methods: the event plane method $\langle v_2 \rangle$ and two-particle correlation method$v_2\{2\}$ \cite{Adams:2004bi, Ollitrault:2009ie}.
One can see that the centrality dependence of $v_2$ can be nicely reproduced by hydrodynamics calculation, and the presence of QGP shear viscosity suppresses the development of $v_2$.
In the figure, both theoretical curves and experimental data for $v_2$ are normalized by $\epsilon_2$ to narrow down the sensitivity to the different experimental methods used for measuring $v_2$.
Since the magnitudes of $\epsilon_2$ differ by about 20\% in two initial condition models, the experimental data are shifted in two panels.
Also due to the choice of different initial condition models, the extracted value of QGP $\eta/s$ has an uncertainty about a factor of $2$-$2.5$.

\subsection{Initial state density and geometry fluctuations}

As has been shown, the knowledge of initial conditions, especially the geometry of the collision zone, is essential for the quantitative extraction of the shear viscosity of the QGP matter produced in heavy-ion collisions.
For many years, most of the dynamical descriptions based on realistic hydrodynamics utilized averaged, smooth and symmetric initial density profiles (see Fig. \ref{fig_e2_v2}). However, quantum fluctuations exist in the earliest stages of the collisions, such as the fluctuations of nucleon positions in nuclei, the fluctuation of color charges in nucleon, and so on.
This leads to lumpy and asymmetric density profiles for the produced fireball which fluctuate one event to another, even for the collisions at a fixed impact parameter \cite{Gyulassy:1996br,Aguiar:2001ac,Andrade:2006yh,Broniowski:2007ft,Andrade:2008xh,Hirano:2009ah,Alver:2010gr,Alver:2010dn,Petersen:2010cw,Staig:2010pn, Qin:2010pf,Ma:2010dv,Xu:2010du,Teaney:2010vd,Qiu:2011iv,Bhalerao:2011yg,Floerchinger:2011qf,Pang:2012he,Schenke:2012wb,Coleman-Smith:2013rla,Luzum:2013yya}.

To quantify the geometry (the geometric anisotropies) of the fluctuating initial states, for each event one may define the eccentricities, $\mathbf{\epsilon}_n = \epsilon_n e^{in\Psi_n}$, where $\epsilon_n$ and $\Phi_n$ are the magnitude and the orientation direction for the $n$-th order eccentricity vector. They may be calculated from the initial density profiles as follows:
\begin{eqnarray}
\mathbf{\epsilon}_n = \epsilon_n e^{in\Phi_n} = \frac{\langle r_\perp^m e^{in\phi}\rangle }{\langle r_\perp^m\rangle } = \frac{\int  r_\perp^m e^{in\phi} \rho(\mathbf{r}_\perp) d^2\mathbf{r}_\perp }{\int r_\perp^m \rho(\mathbf{r}_\perp)  d^2\mathbf{r}_\perp } \,,
\end{eqnarray}
where $\rho(\mathbf{r}_\perp)$ is the (energy or entropy) density distribution in the transverse plane.
Here the power index $m$ is the radial weight, and often taken to be $m=n$ for $n \ge 2$ \cite{Petersen:2010cw,Staig:2010pn,Qin:2010pf}.
For $n=1$ (dipole asymmetry), it is suggested to take $m=3$ \cite{Teaney:2010vd}.
One can see that for smooth initial conditions without fluctuations as shown in Fig. \ref{fig_e2_v2}, all odd $\epsilon_n$ coefficients vanish.

\begin{figure}
\begin{center}
\includegraphics*[width=12.0cm, height=2.5cm]{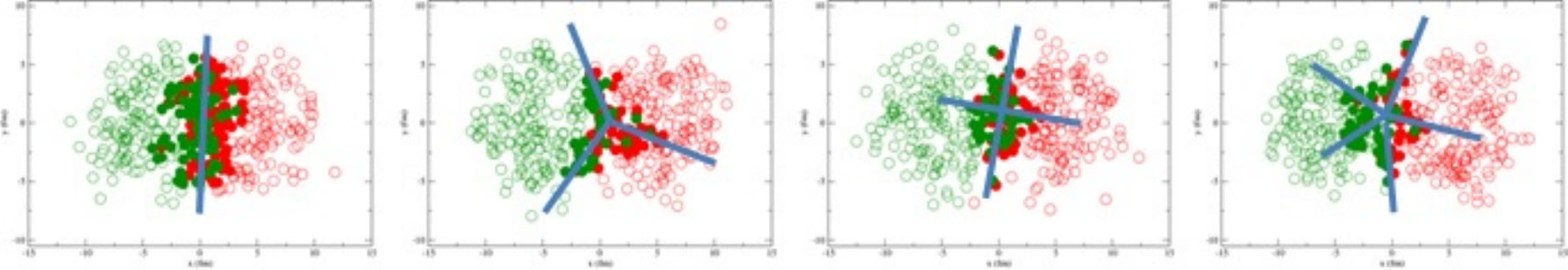}
\caption{(Color online) The simulation of the fluctuating initial states in Au-Au collisions at $200$~AGeV at RHIC using a Glauber Monte-Carlo model \cite{Qin:2010pf}. These events are selected on purpose to obtain large values of eccentricities: $\epsilon_2$, $\epsilon_3$, $\epsilon_4$ and $\epsilon_5$ (from left to right).
}
\label{fig_qin_ic}
\end{center}
\end{figure}

\begin{figure}
\begin{center}
\includegraphics*[width=9.0cm, height=6.0cm]{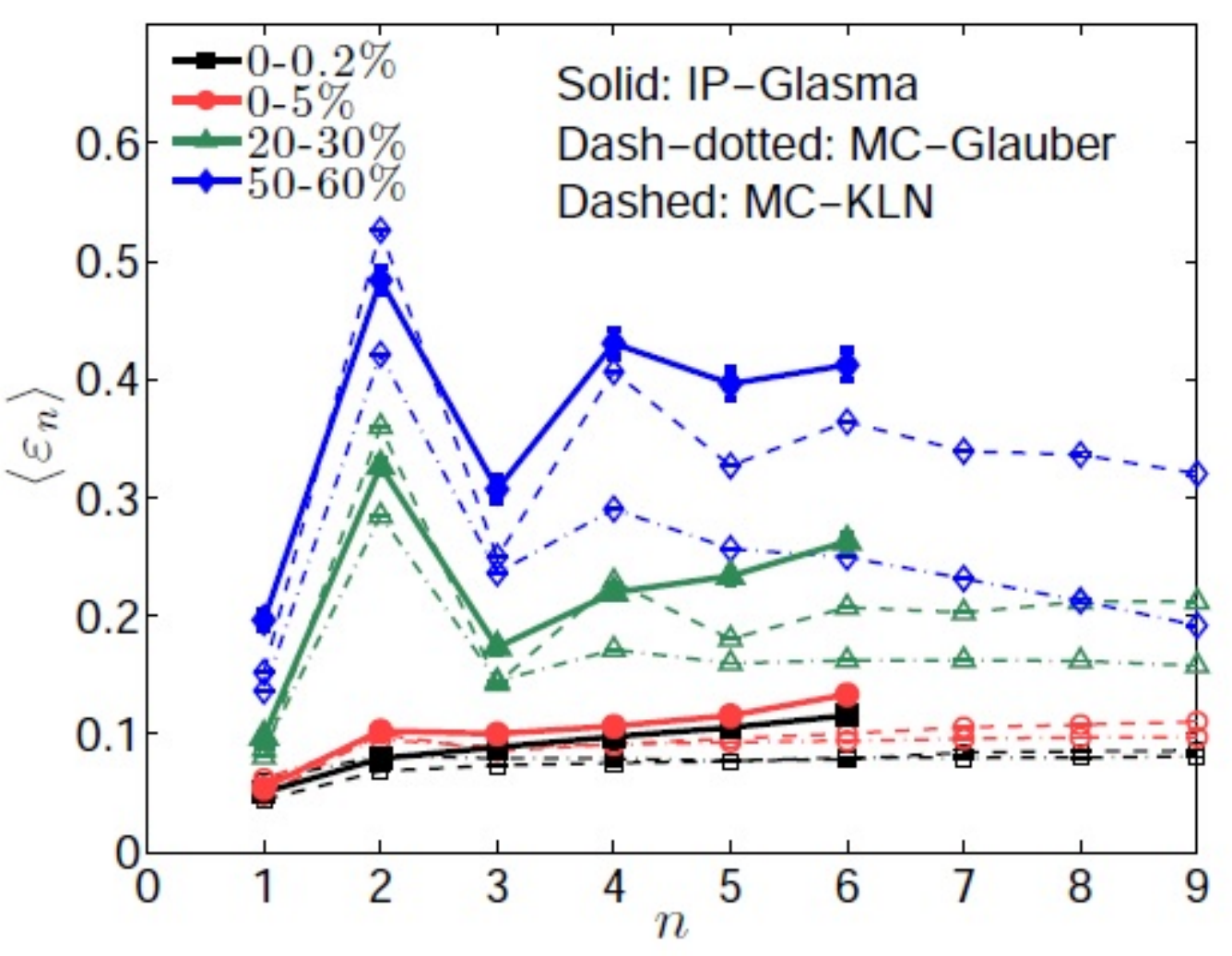}
\caption{(Color online) The power spectrum of the initial state eccentricities $\epsilon_n$ for Pb-Pb collisions at $2.76$~ATeV.
Results are shown for four collision centralities and three initial condition models (MC-Glauber, MC-KLN and IP-Glasma).
The figure is taken from Ref. \cite{Heinz:2013th}.
}
\label{fig_heinz_en_spectrum}
\end{center}
\end{figure}

Fig. \ref{fig_qin_ic} shows a simulation of the fluctuating initial states produced in Au-Au collisions at $200$~AGeV at RHIC using a Monte-Carlo Glauber model \cite{Qin:2010pf}.
One can see that the density profiles of the initial state collision zones are lumpy and anisotropic in the transverse plane, and the initial conditions fluctuate violently from one event to another.
The events shown in the figure have been selected on purpose in order to obtain large values for $\epsilon_2$, $\epsilon_3$, $\epsilon_4$ and $\epsilon_5$ (from left to right), so the initial collision zones are mainly dominated by the elliptic, triangular, quadrangular, and pentagonal shapes, respectively.

Fig. \ref{fig_heinz_en_spectrum} shows the initial state $\epsilon_n$ power spectrum for Pb-Pb collisions at $2.76$~ATeV at the LHC.
Results are shown for four collision centralities, and three initial condition models (MC-Glauber \cite{Miller:2003kd}, MC-KLN \cite{Drescher:2006ca} and IP-Glasma \cite{Schenke:2012wb}).
One can see that in the ultra-central ($0-0.2\%$) collisions, all $\epsilon_n$'s have roughly the same magnitudes since they all originate from the initial state fluctuations.
For less central collisions, since the collision zone has a pronounced elliptic shape, the values of $\epsilon_2$ (and to some degree $\epsilon_4$, $\epsilon_6$ as well) are larger than the odd $\epsilon_n$ coefficients.

\subsection{Hydrodynamic response and high-order anisotropic flow}

As a consequence of the interaction among the medium constituents, hydrodynamic evolution of the fireball will translate the initial geometric eccentricity into final state momentum anisotropy. To quantify such effect, one may perform the Fourier decomposition for the transverse momentum distribution of final state particles,
\begin{eqnarray}
\frac{dN}{dydp_Td\psi} \propto 1 + 2 \sum_n v_n(p_T, y) \cos\left[n(\psi - \Psi_n(p_T, y))\right] \,,
\end{eqnarray}
where $v_n$ and $\Psi_n$ are the magnitude and orientation angle (event plane) of the $n$-th order anisotropic flow vector $\mathbf{v}_n = v_n e^{in\Psi_n}$.
Note that $v_n$ and $\Psi_n$ are defined for a single collision, and can be $p_T$ (and $y$) dependent or integrated.
The anisotropic flow may be obtained from the final state momentum distribution as follows:
\begin{eqnarray}
\mathbf{v}_n = v_n e^{in\Psi_n} = \langle  e^{in\psi} \rangle
\end{eqnarray}

\begin{figure}
\begin{center}
\includegraphics*[width=6.0cm]{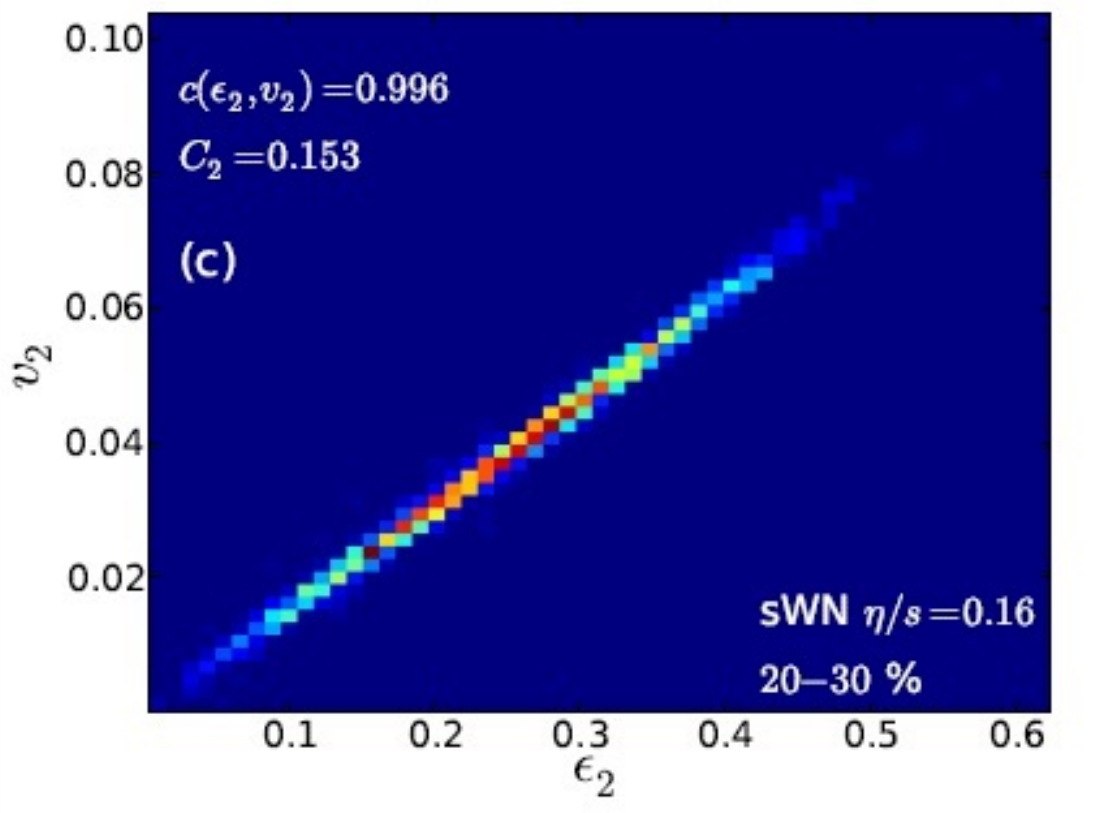}
\includegraphics*[width=6.0cm]{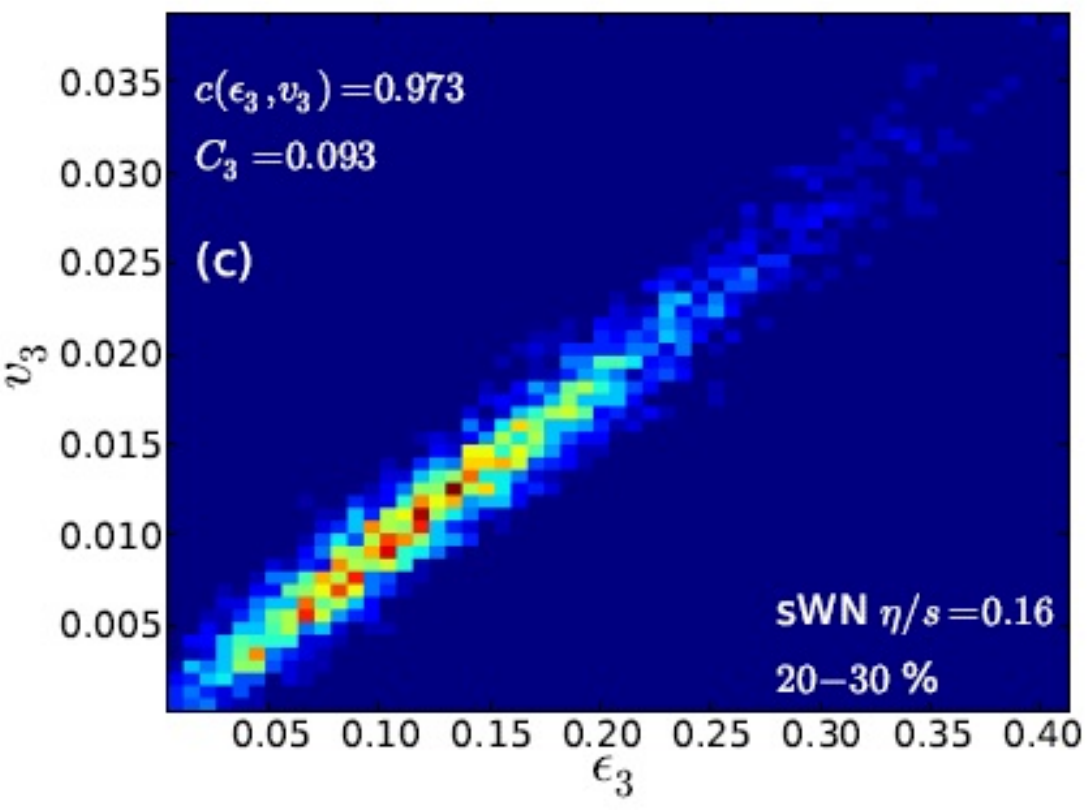}
\caption{(Color online) Scattered plots for $(\epsilon_2, v_2$) (left panel) and $(\epsilon_3, v_3)$ (right panel) from event-by-event hydrodynamics calculations for 20-30\% central Au-Au collisions at $200$~AGeV at RHIC (from Ref. \cite{Niemi:2012aj}).
}
\label{fig_niemi_v23_vs_e23}
\end{center}
\end{figure}

\begin{figure}
\begin{center}
\includegraphics*[width=9.0cm, height=6.0cm]{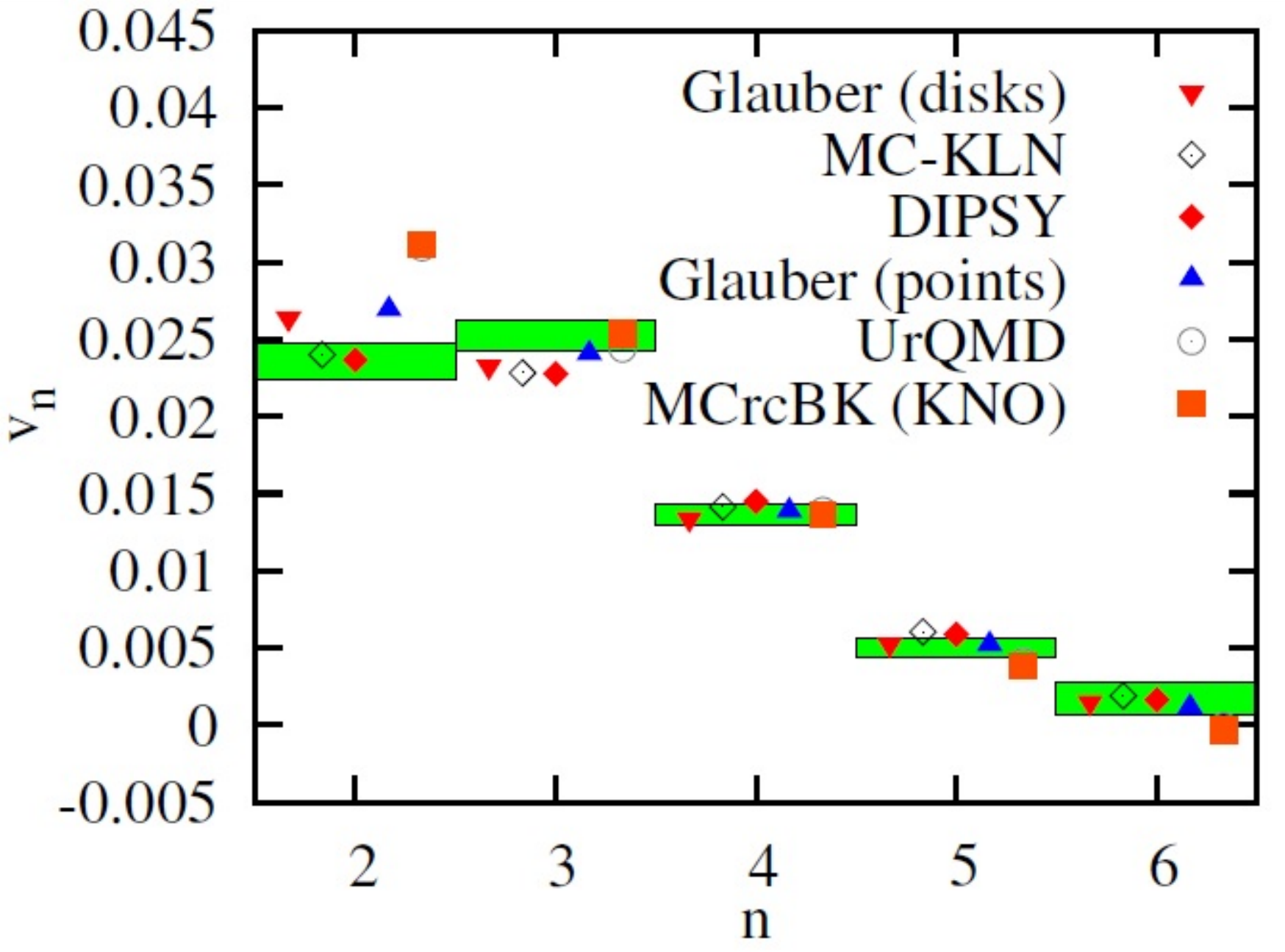}
\caption{(Color online) The $v_n$ power spectrum in ultra-central (0-0.1\% centrality) Pb-Pb collisions at $2.76$~AGeV at the LHC \cite{ATLAS:2012at} compared to viscous hydrodynamics calculations utilizing a few different initial condition models (from Ref. \cite{Luzum:2012wu}).
}
\label{fig_luzum_vn_spectrum}
\end{center}
\end{figure}

Fig. \ref{fig_niemi_v23_vs_e23} show the scattered plots for $(\epsilon_2, v_2$) (left panel) and $(\epsilon_3, v_3)$ (right panel) from event-by-event hydrodynamics calculations for 20-30\% central Au-Au collisions at $200$~AGeV at RHIC \cite{Niemi:2012aj}.
One can see that the anisotropic flow coefficients $v_2$ ($v_3$) show strong correlation to the initial state eccentricities $\epsilon_2$ ($\epsilon_3$).
This is true for their orientation angles as well: the final state event plane $\Psi_n$ are strongly correlated to the initial participant plane $\Phi_n$ for $n=2$ and $n=3$ \cite{Petersen:2010cw, Qin:2010pf, Gardim:2011qn}.
The linear response $v_2/\epsilon_2$ is stronger than $v_3/\epsilon_3$.
For higher order harmonics, the linear correlation between $\epsilon_n$ and $v_n$ is spoiled by the non-linear contribution from lower-order harmonics \cite{Qiu:2011iv}.

Fig. \ref{fig_luzum_vn_spectrum} show the final state $v_n$ power spectrum for ultra-central (0-0.1\% centrality) Pb-Pb collisions at $2.76$~AGeV at the LHC compared to the viscous hydrodynamics calculations using a few different initial condition models \cite{Luzum:2012wu}.
One can see that in hydrodynamics calculations, higher order anisotropic flow harmonics are typically suppressed compared to lower order harmonics.
This is true even in ideal hydrodynamics.
We remind that all the initial eccentricities have roughly the same magnitudes in ultra-central collisions as shown in Fig. \ref{fig_heinz_en_spectrum}.
This is understandable since hydrodynamic evolution is more sensitive to larger scale structure (lower momentum modes) of the initial density profiles, and less sensitive to smaller scale structure (thus higher order harmonics are suppressed).

\begin{figure}
\begin{center}
\includegraphics*[width=6.25cm, height=5.0cm]{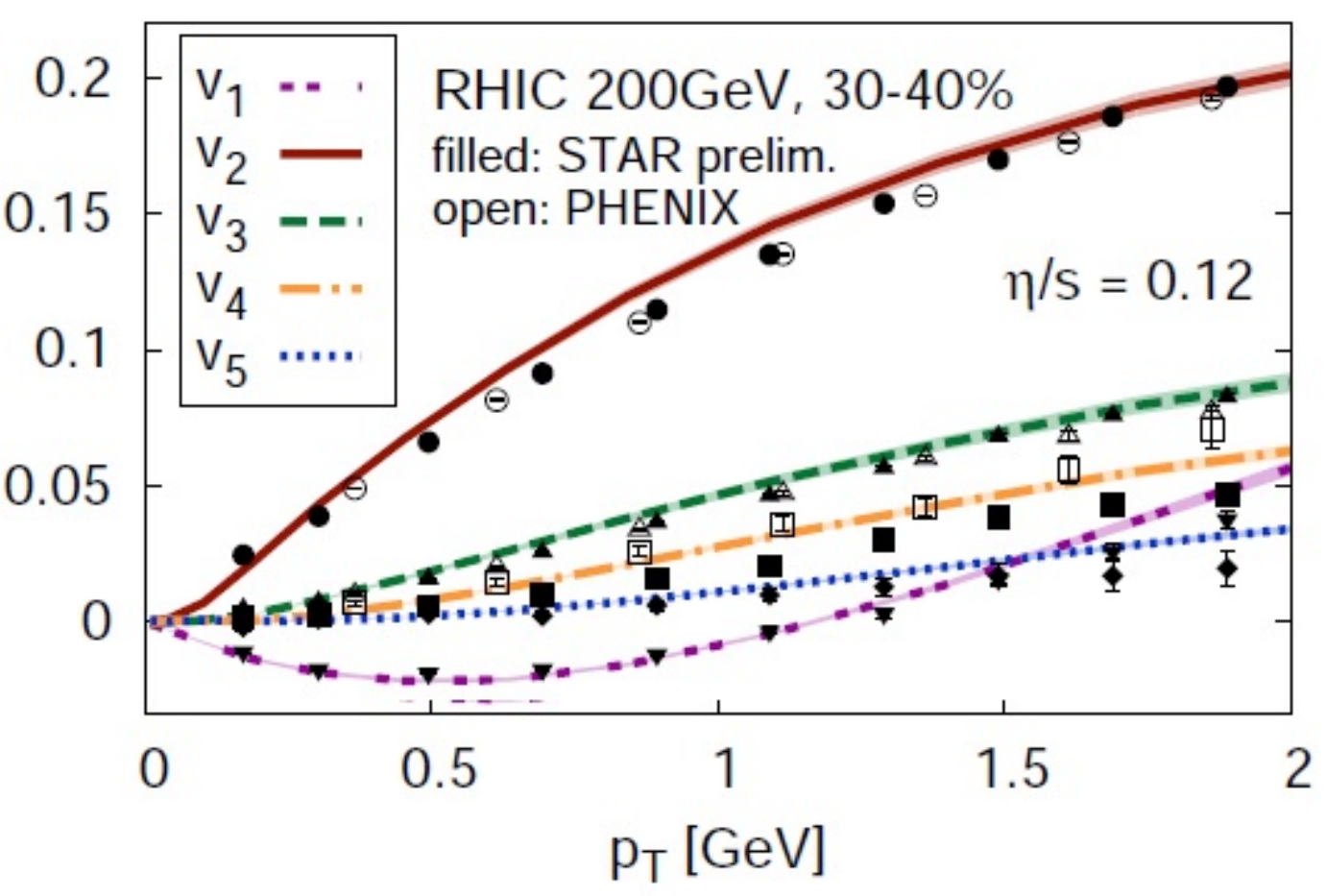}
\includegraphics*[width=6.25cm, height=5.0cm]{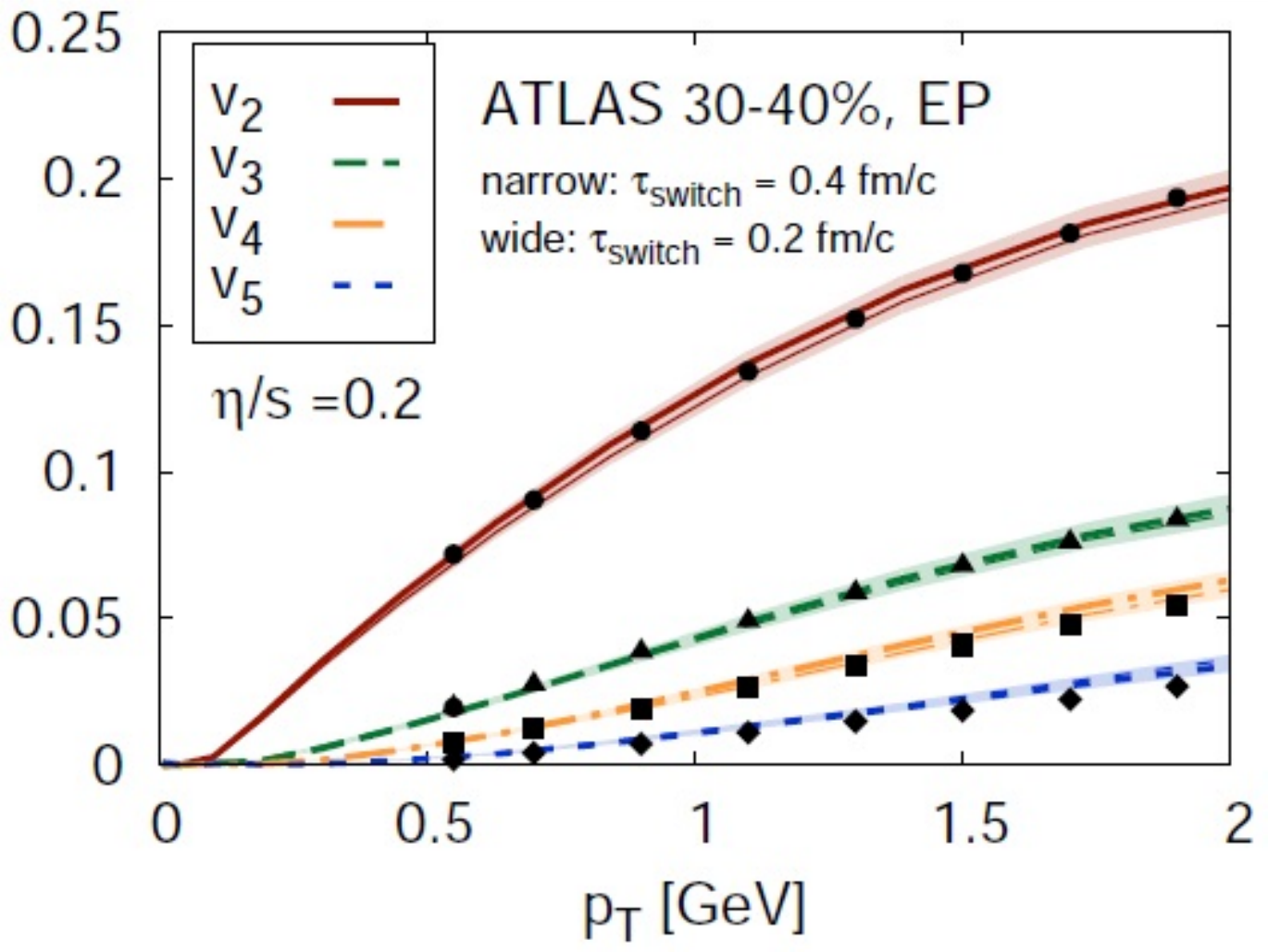}
\caption{(Color online) The anisotropic flow coefficients $v_n$ as a function of $p_T$ in Au-Au collisions at $200$~AGeV at RHIC (left panel) and in Pb-Pb collisions at $2.76$~AGeV at the LHC (right panel) (from Ref. \cite{Schenke:2010rr}).
}
\label{fig_schenke_vn_pt}
\end{center}
\end{figure}

Fig. \ref{fig_schenke_vn_pt} shows the anisotropic flow coefficients $v_n$ as a function of $p_T$ in Au-Au collisions at $200$~AGeV at RHIC (left panel) and in Pb-Pb collisions at $2.76$~AGeV at the LHC (right panel). The theoretical results are from (3+1)-dimensional viscous hydrodynamics calculations using the IP-Glasma+MUSIC model \cite{Schenke:2010rr}.
The data at RHIC are from PHENIX \cite{Adare:2011tg} and STAR \cite{Pandit:2012mq}. The data points at the LHC are from ATLAS \cite{ATLAS:2012at}.
The best descriptions to the experimental data give the average value of $\eta/s$ to be $0.12$ at RHIC and $0.2$ at the LHC.
This means that on average the QGP medium produced at the LHC is less strongly coupled than that at RHIC.
Since the temperature of the medium is higher at the LHC, this suggests that there is a strong temperature dependence for $\eta/s$.
The precise determination of the temperature-dependent specific shear viscosity $\eta/s (T)$ is one of the essential tasks in the current study of heavy-ion collisions.
We note that with the use of the parameterization of temperature-dependent $\eta/s(T)$ from Ref. \cite{Niemi:2011ix},  a reasonable description of the flow data was also obtained from the IP-Glasma+MUSIC model \cite{Schenke:2010rr}.

\subsection{Flow fluctuations and correlations}

\begin{figure}
\begin{center}
\includegraphics*[width=12.0cm]{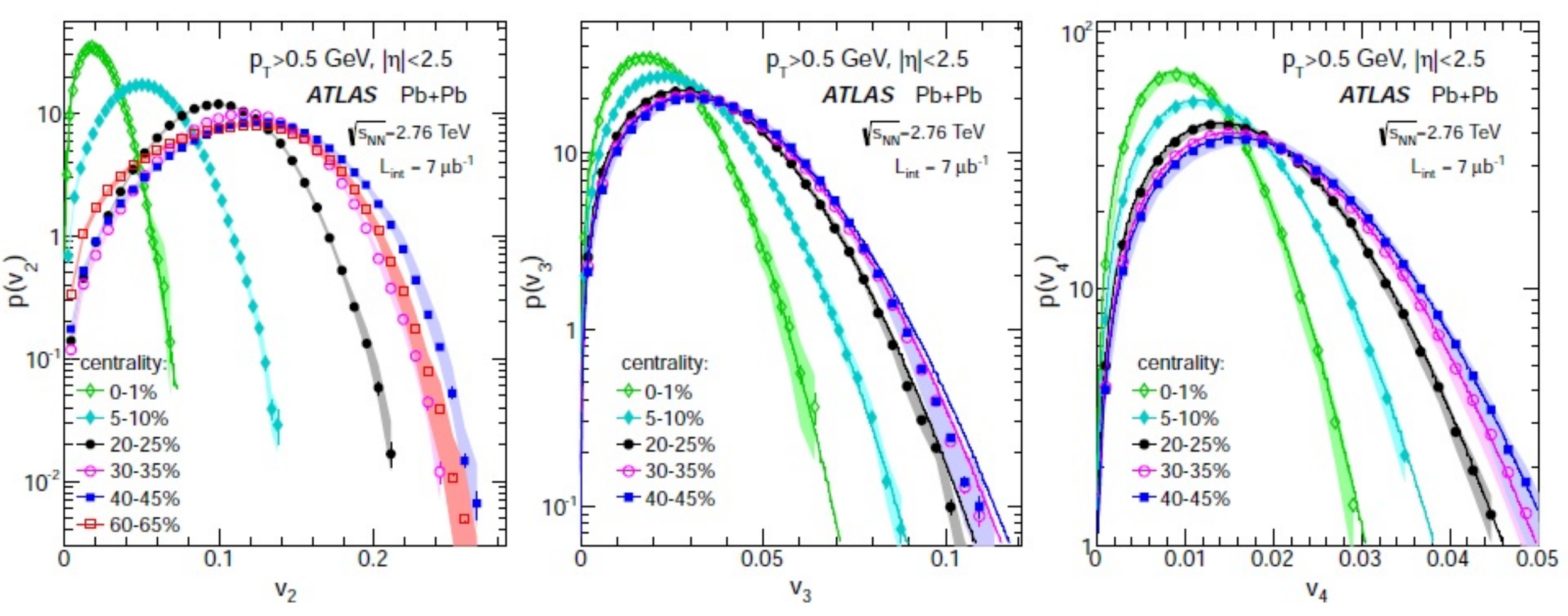}
\caption{(Color online) The event-by-event $v_2$ (left panel), $v_3$ (middle panel) and $v_4$ (right panel) distributions for charged particles with $p_T > 0.5$~GeV in several centrality classes for Pb-Pb collisions at $2.76$~TeV at the LHC, measured by ATLAS \cite{Aad:2013xma}. %Curves are fits to Bessel-Gaussian function.
}
\label{fig_atlas_vn_distribution}
\end{center}
\end{figure}

Due to large fluctuations in the initial states, both $\epsilon_n$ and $v_n$ vary strongly from one event to another, even for a very narrow centrality bin.
Since the averaged $v_n$ mainly reflect the hydrodynamic response to the averaged initial collision geometry of the produced fireball, the measurements of event-by-vent $v_n$ distributions is essential in order to obtain direct insights into the fluctuations in the initial states.
Both ATLAS and ALICE Collaborations have measured the event-by-event $v_n$ distributions for Pb-Pb collisions at $2.76$~ATeV at the LHC \cite{Aad:2013xma, Timmins:2013hq}.
Fig. \ref{fig_atlas_vn_distribution} shows the results from ATLAS in several centrality bins.
One feature of $v_n$ distributions is that they become broader from more central to more peripheral collisions due to the increasing of the $v_n$ magnitudes.
A strong centrality dependence is observed for the shape of $v_2$ distribution, while higher-order harmonics shows much smaller centrality dependence.
The event-by-event $v_n$ distributions have also been studied in some details by a few groups utilizing relativistic hydrodynamics with various initial condition models \cite{Gale:2012rq, Niemi:2012aj, Renk:2014jja}.

\begin{figure}
\begin{center}
\includegraphics*[width=12.0cm]{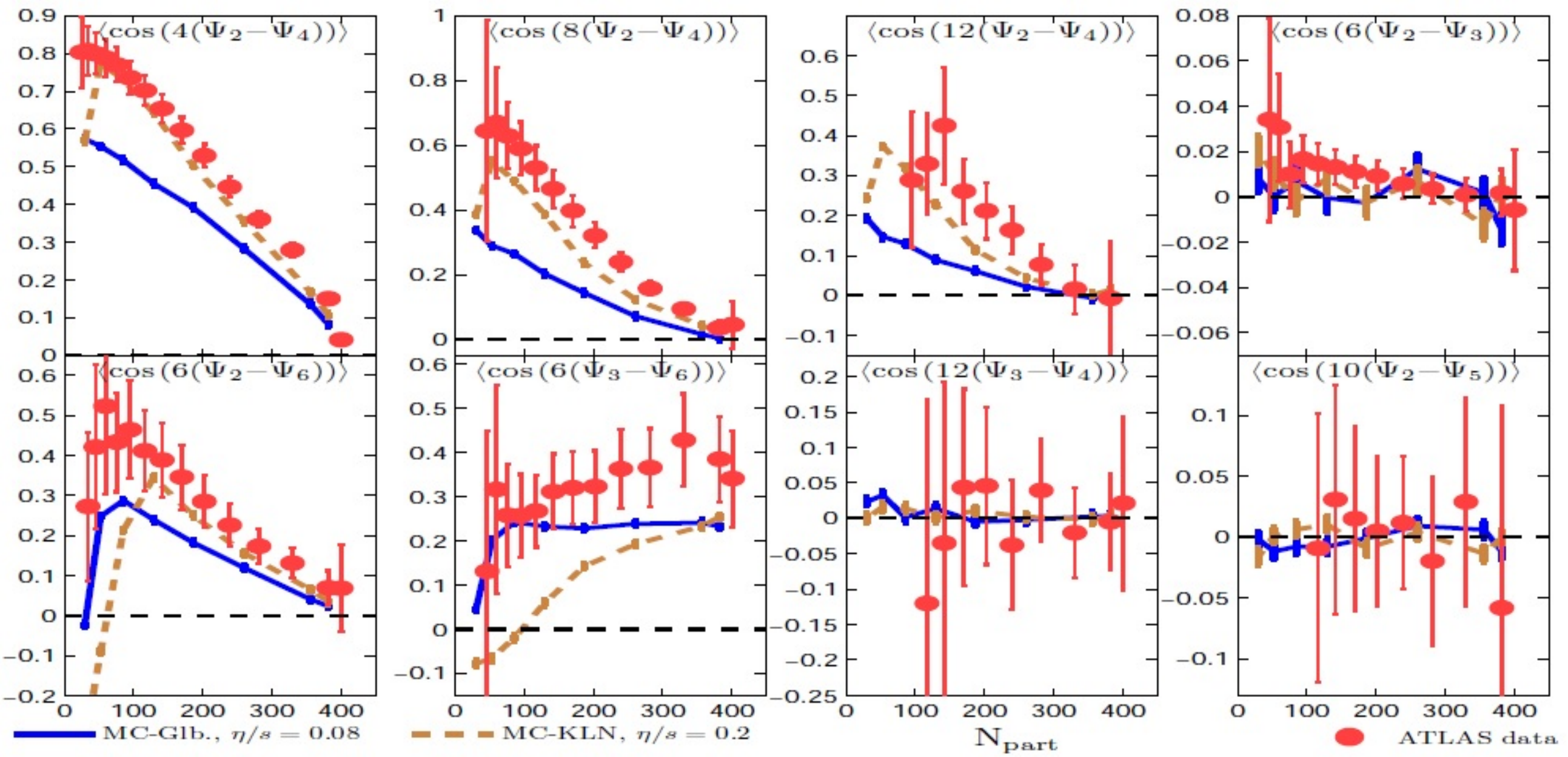}
\caption{(Color online) Correlations between two different event planes $\langle \cos[jk(\Psi_m - \Psi_n)] \rangle$, where $j$ is an integer, and $k$ is the least common multiple of $m$ and $n$ (from Ref. \cite{Qiu:2012uy}).
}
\label{fig_qiu_phin_correlations}
\end{center}
\end{figure}

\begin{figure}
\begin{center}
\includegraphics*[width=12.0cm]{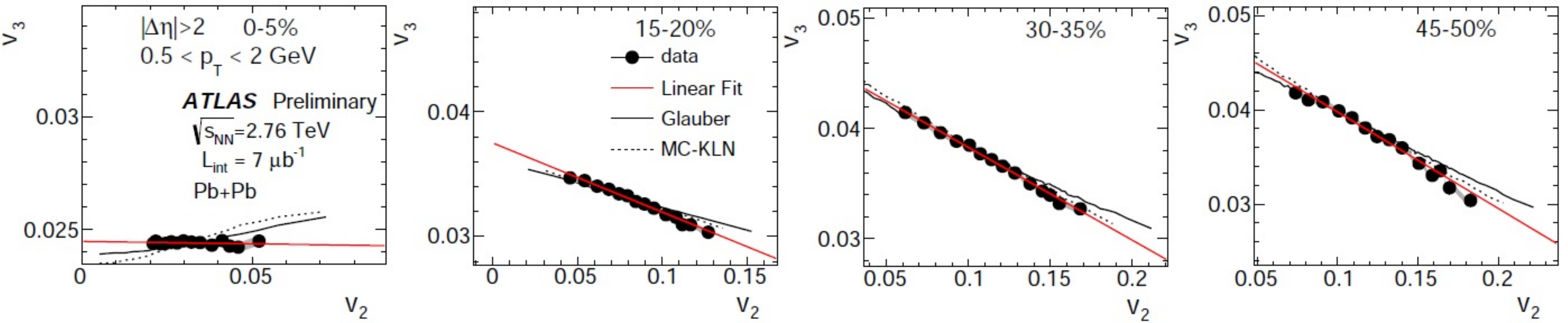}
\includegraphics*[width=12.0cm]{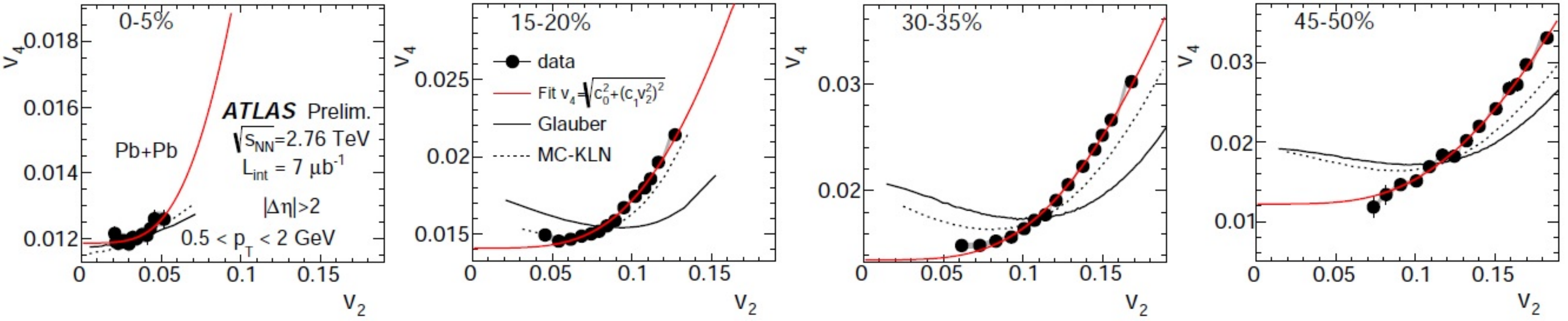}
\caption{(Color online) The $v_3(v_2)$ (top panels) and $v_4(v_2)$ (bottom panels) correlations measured in $0.5 < p_T < 2$~GeV in three centrality bins.
In each panel, the correlation data are fit to functions that include both linear and non-linear contributions.
The correlation data are also compared with re-scaled $\epsilon_n(\epsilon_2)$ correlation from the MC Glauber and MC-KLN models in the same centrality bins.
The figures are taken from Ref. \cite{ATLAS-CONF-2014-022}.
}
\label{fig_atlas_vn_vs_v2}
\end{center}
\end{figure}

Various hydrodynamics calculations have shown that the elliptic flow $v_2$ and the triangular flow $v_3$ are mainly driven by initial state $\epsilon_2$ and $\epsilon_3$.
However, higher-order flow harmonics $v_n$ ($n > 3$) arise from a combination of $\epsilon_n$ and the non-linear mixing with lower-order harmonics.
Such non-linear mode-mixing will manifest in the final state correlations between different order flow harmonics.
Fig. \ref{fig_qiu_phin_correlations} shows the correlations between two different event-plane angles as a function of collision centrality.
Experimental data are from Ref. \cite{Jia:2012sa} and theoretical curves are from Ref. \cite{Qiu:2012uy} with the use of two different initial condition models, MC-Glauber \cite{Miller:2003kd} and MC-KLN \cite{Drescher:2006ca} (for later hydrodynamics evolution, $\eta/s$ is taken to be $0.08$ and $0.2$ for these two models, respectively).
One can see that hydrodynamic calculation can qualitatively describe the observed centrality dependence of event plane correlations.
We note that several final state event plane correlators exhibit very different centrality dependence from the initial state participant plane correlators \cite{Bhalerao:2011bp, Qin:2011uw, Jia:2012ju, Qiu:2012uy}. This is mainly due to the development of the mode-mixing between different flow harmonics as a result of non-linear hydrodynamics evolution of the fireball.

Besides the correlations between different flow angles $\Psi_n$, the magnitudes of different flow coefficients $v_n$ correlate to each other as well.
Fig. \ref{fig_atlas_vn_vs_v2} shows the ATLAS measurement of the correlations between the magnitudes of different order flow harmonics ($v_3$-$v_2$ and $v_4$-$v_2$ correlations) using the event-shape selection method \cite{ATLAS-CONF-2014-022}.
One can see that $v_3$ is anti-correlated to $v_2$, which is very similar to the correlation between $\epsilon_3$ and $\epsilon_2$ in the initial states.
This indicates $v_2$-$v_3$ correlation reflect mostly initial geometry effects; this is expected since $v_2$ and $v_3$ are dominated by the linear hydrodynamics response.
The $v_2$-$v_3$ correlation data are fitted with a linear function $v_3 = kv_2 + v_3^{0}$, one can see that the magnitude of $|k|$, which characterizes the strength of the anti-correlation, increases from central to peripheral collisions.
This anti-correlation is understandable since when the overlap region becomes more elliptic, the fluctuation to large triangularity is constrained.
For $v_2$-$v_4$ correlation, both linear and non-linear collective dynamics contribute.
The correlation data are well described by two-parameter fits with the form: $v_4^2 = c_0^2 + (c_1 v_2^2)^2$, where the linear term $v_4^L = c_0$ is associated with $\epsilon_4$ and the non-linear contribution $v_4^{NL} = c_1 v_2^2$ reflects the $v_2$-$v_4$ mixing.
The linear term depends weakly on centrality, whereas the non-linear term increases as the collisions become more peripheral \cite{ATLAS-CONF-2014-022} (not shown).

\subsection{Longitudinal fluctuations}

Whe studying the anisotropic flow harmonics and final state correlations, one often utilizes a large pseudo-rapidity gap between correlated particles in order to minimize the contribution from non-flow effects such as resonance decays and jets.
The use of large rapidity gap is reasonable when the initial density distribution and final state flow harmonics at different rapidities are prefectly correlated.
However, the density profiles of initial states fluctuate not only in the transverse plane, but also in the longitudinal direction.
Note that the fluctuations due to finite multiplicity for a given event are usually corrected using the sub-event method \cite{Poskanzer:1998yz,Ollitrault:1993ba}.
Initial state longitudinal fluctuations may lead to the fluctuations and decorrelations of the final flow orientations at different pseudo-rapidities \cite{Petersen:2011fp, Xiao:2012uw, Jia:2014ysa, Jia:2014vja, Pang:2014pxa}.
One consequence is the reduction of the values of the final state anisotropic flows \cite{Pang:2012he}.

\begin{figure}
\begin{center}
\includegraphics*[width=12.0cm]{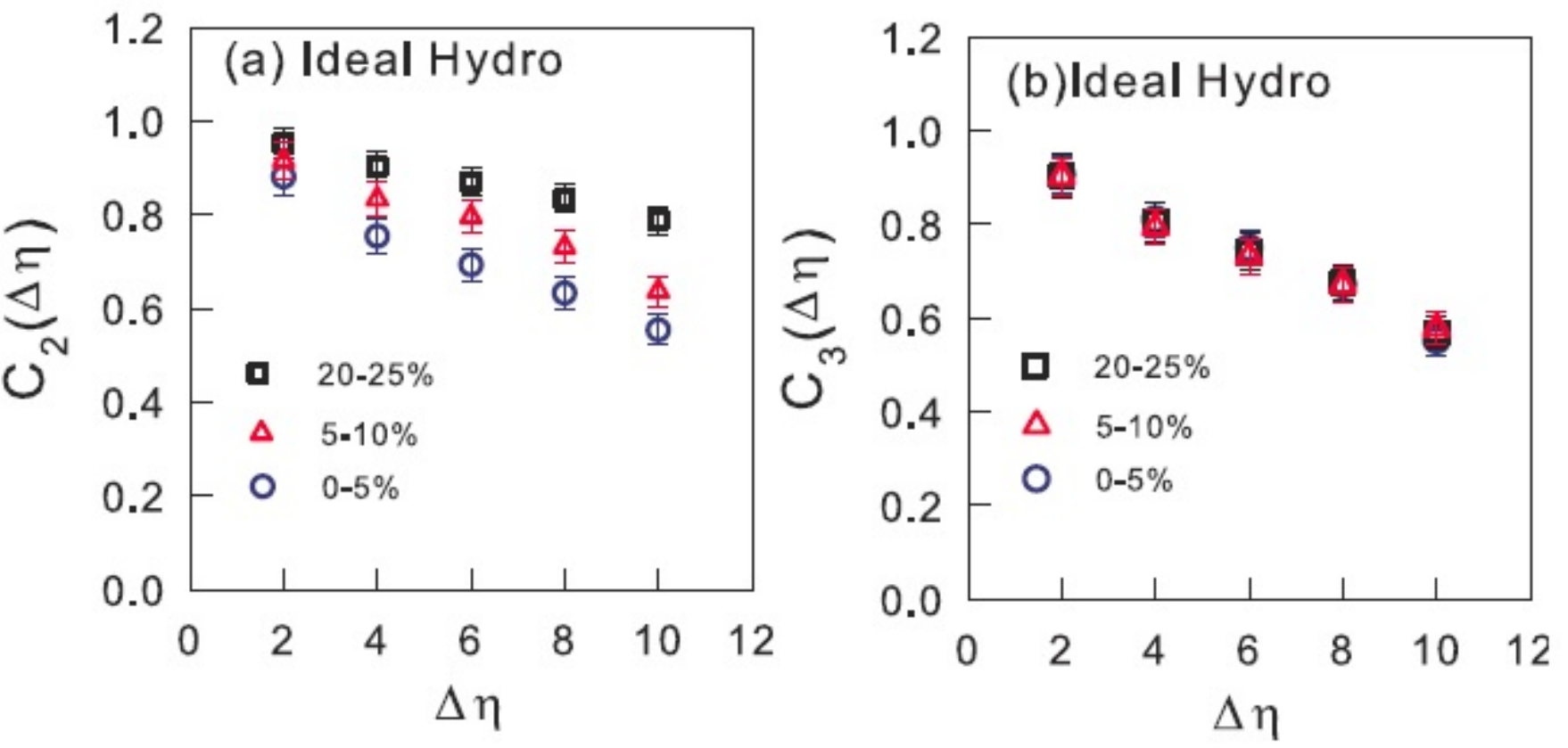}
\caption{(Color online) The longitudinal correlation functions $C_2(\Delta \eta)$ (left panel) and $C_3(\Delta\eta)$ (right panel) in Pb-Pb collisions at $2.76$~AGeV for three different centrality bins (from Ref. \cite{Pang:2014pxa}).
}
\label{fig_longitudinal_decorrelation}
\end{center}
\end{figure}

To study the effect of longitudinal fluctuations, one may define the following correlation function between two rapidity bins $A$ and $B$ \cite{Pang:2014pxa}:
\begin{eqnarray}
C_n(A, B) =  \frac{\langle \mathbf{Q}_{n}(A) \mathbf{Q}_{n}^*(B) \rangle }{\sqrt{\langle \mathbf{Q}_{n}(A) \mathbf{Q}_{n}^*(A) \rangle}\sqrt{ \langle \mathbf{Q}_{n}(B) \mathbf{Q}_{n}^*(B) \rangle}}\,,
\label{DefCn_hydro}
\end{eqnarray}
where $\mathbf{Q}_n = Q_n e^{in\Psi_n} = \langle e^{in\phi} \rangle$. In the continuum limit, $\mathbf{Q}$ is identical to flow vector $\mathbf{v}_n$.
Fig. \ref{fig_longitudinal_decorrelation} shows the correlation functions $C_2$ (left panel) and $C_3$ (right panel) from relativistic hydrodynamics calculation for three different centrality bins for Pb-Pb collisions at the LHC \cite{Pang:2014pxa}.
One can see that the anisotropic flows at different pseudo-rapidities are not perfectly correlated, and the degree of such decorrelation becomes stronger with increasing pseudo-rapidity gap.
The correlation $C_2$ of the second-order anisotropic flow shows a strong centrality dependence whereas the correlation $C_3$ of the third-order anisotropic flow is independent of centrality.
This can be understood since $v_3$ arises purely from initial state fluctuations which is almost independent of collision geometry, while $v_2$ has a strong dependence on the collision centrality (the geometry of the collision zone).
It was further shown that the longitudinal decorrelatons in the anisotropic flows are caused by the longitudinal fluctuations in the initial state density distributions with a twist structure as well as as additional random fluctuations on top of a twist \cite{Pang:2014pxa}.

\subsection{Anisotropic flow in small collision systems}

The relativistic proton-proton (p-p), proton-nucleus (p-A) and deuteron-nucleus (d-A) collisions are expected to provide the baseline for studying the transport properties of the hot and dense QCD matter produced in A-A collisions.
The collective behavior and anisotropic flow are expected to be much weaker in these colliding systems since the size of the produced matter is much smaller (even though a mini quark-gluon plasma is produced).
However, ALICE, ATLAS and CMS Collaborations have observed a clear collective behavior in p-Pb collisions at $5.02$~ATeV \cite{Abelev:2012ola, Aad:2012gla, Chatrchyan:2013nka}, very similar to peripheral Pb-Pb collisions at $2.76$~ATeV at the LHC.
Similar correlation results have been obtained for d-Au collision at RHIC \cite{Adare:2013piz} and high multiplicity p-p events at the LHC energies \cite{Khachatryan:2010gv}.

\begin{figure}
\begin{center}
\includegraphics*[width=9.0cm, height=6.0cm]{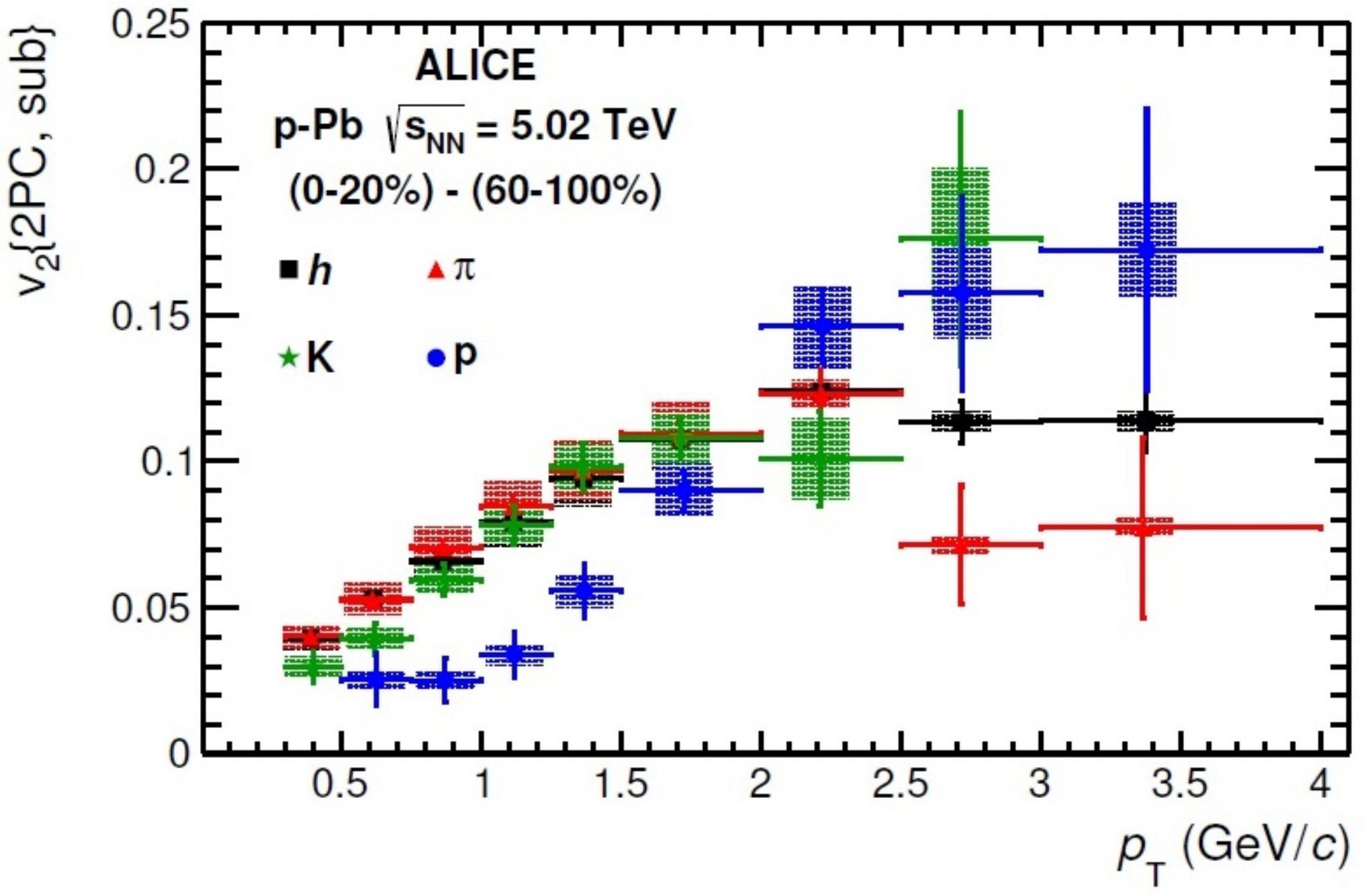}
\caption{(Color online) The Fourier coefficient $v_2\{2\}$ for hadrons, pions, kaons and protons as a function of $p_T$ measured by ALICE Collaboration using two-particle correlations in the 0-20\% multiplicity class after the subtraction of that from the 60-100\% multiplicity class (from Ref. \cite{ABELEV:2013wsa}).
}
\label{fig_pPb_vn}
\end{center}
\end{figure}

The origins of the observed correlations and collective behavior in these smaller colliding systems are still in debate.
One natural explanation for the observed collectivity and correlations is the hydrodynamic expansion of the fireball with fluctuating initial conditions \cite{Bozek:2013uha, Bzdak:2013zma, Qin:2013bha, Schenke:2014zha}, which suggests that a mini-QGP may be produced in these small colliding systems.
A strong support for hydrodynamics explanation is the mass ordering of the elliptic flow $v_2$ observed by the ALICE \cite{ABELEV:2013wsa} (see Fig. \ref{fig_pPb_vn}), which was reproduced later on by hydrodynamic calculations \cite{Bozek:2013ska, Werner:2013ipa}.
Recently Ref. \cite{Ma:2014pva} has shown that the long-range azimuthal correlations in both p-p and p-Pb collisions can be explained by the incoherent scattering of partons using the AMPT model with string melting \cite{Lin:2004en}.
The observed two-particle azimuthal correlations was also fitted by the two-gluon emission mechanism in the initial states within the framework of Color Glass Condensate \cite{Dusling:2013oia}.
Recently relativistic $^3$He-Au collisions were proposed in order to exploit the intrinsic triangular geometry in the initial sates \cite{Nagle:2013lja,Bozek:2014cya}.
By detailed and systematic comparisons of relativistic p-A, d-A, $^3$He-A and A-A collisions, it is promising that one can disentangle the initial geometry contribution from the final state viscosity effect.

\section{Color Opacity of QGP}

Hard partonic jets that are produced from early stage scatterings provide important probes to study the properties of the QGP created in high energy nuclear collisions.
During their propagation through the hot and dense nuclear matter, they interact with the medium constituents via elastic and inelastic collisions, and usually lose energy in the process \cite{Bjorken:1982tu,Gyulassy:1990ye,Wang:1991xy}. This is often referred to as jet quenching.
The main purpose of jet quenching studies in heavy-ion collisions is to understand the detailed mechanisms of jet-medium interaction from which one may infer useful information about the produced hot and dense QGP.

\begin{figure}
\begin{center}
\includegraphics*[width=7.2cm]{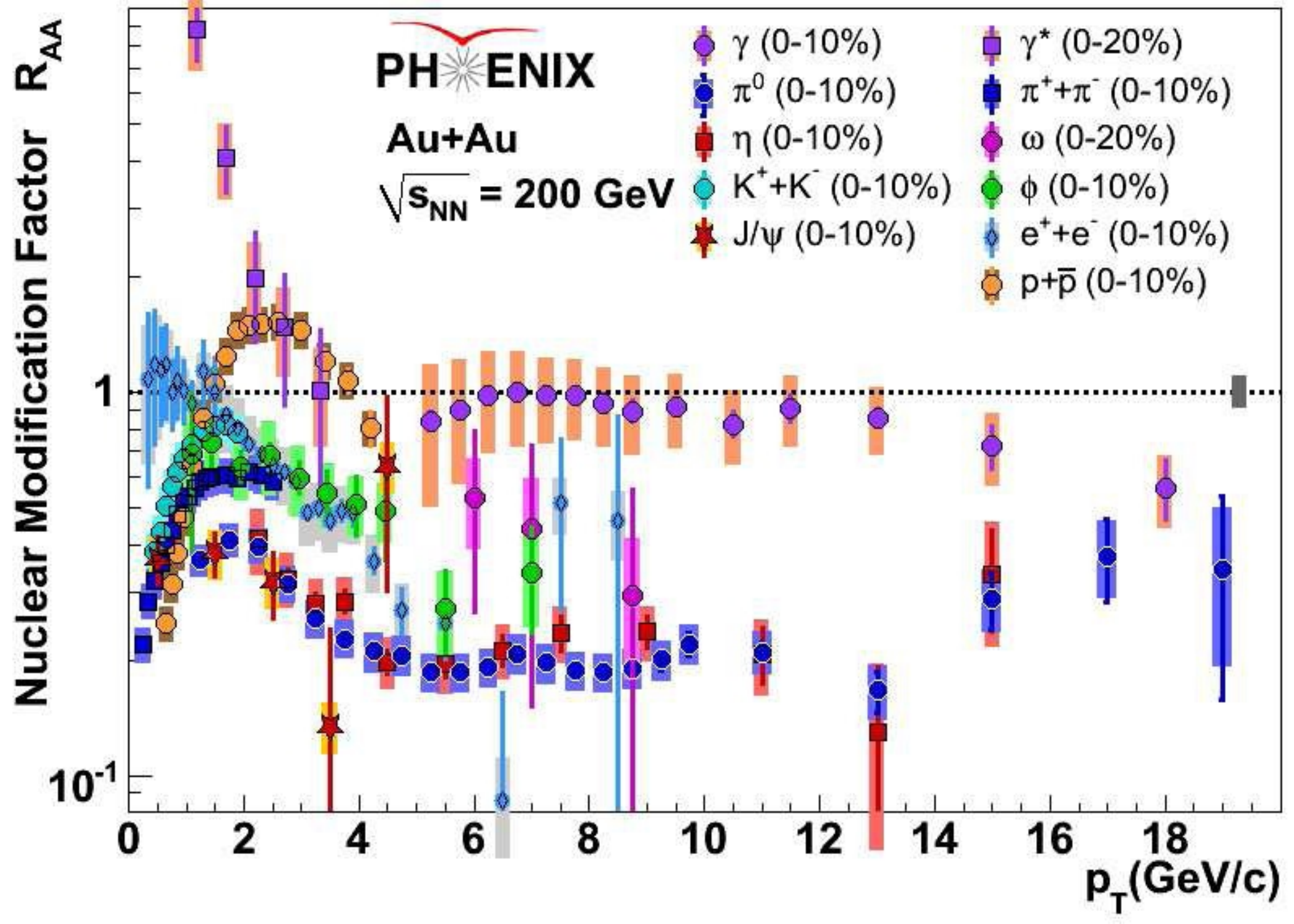}
\includegraphics*[width=5.3cm]{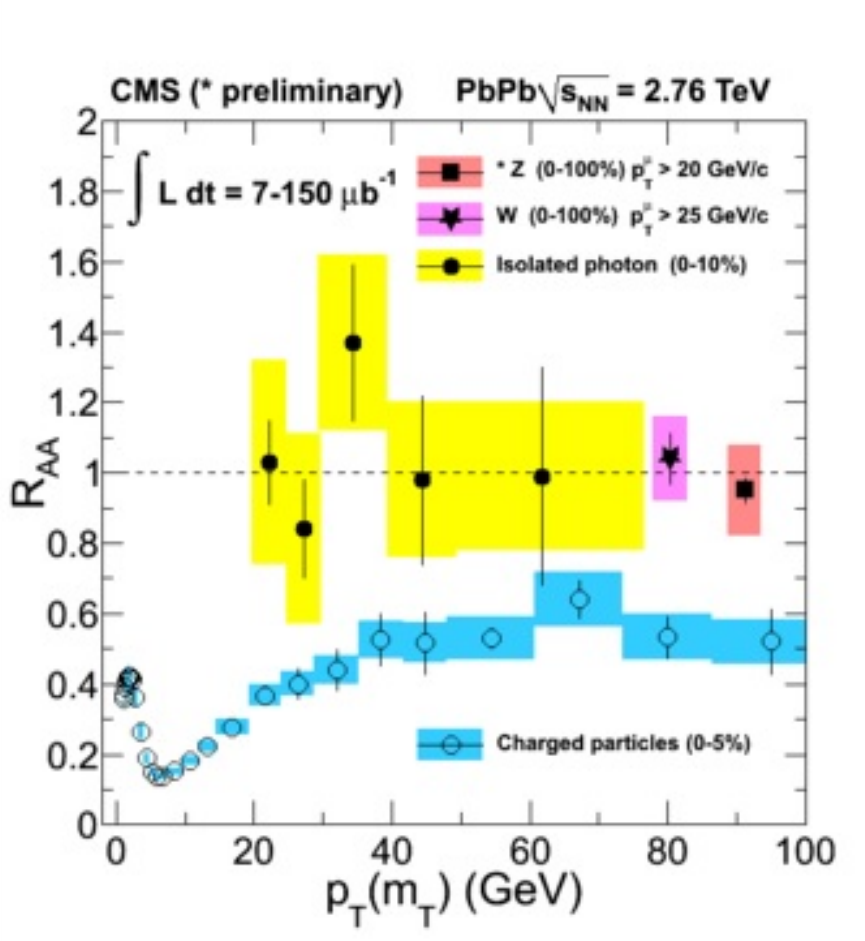}
\caption{(Color online) Left: $R_{AA}$ for hadrons, direct photons and non-photonic electrons in most central Au-Au collisions at $200$~AGeV at RHIC, measured by PHENIX \cite{Sharma:2011zz}. Right: $R_{AA}$ for hadrons, direct photons, and Z/W bosons in most central Pb-Pb collisions at $2.76$~ATeV at the LHC, measured by CMS \cite{CMS:2012aa}.
}
\label{fig_RAA_ID}
\end{center}
\end{figure}

One of the important jet quenching observables is the suppression of the single inclusive hadron yield at high transverse ($p_T$) in A-A collisions as compared to that from elementary p-p collisions  \cite{Adcox:2001jp,Adler:2002xw,Aamodt:2010jd}.
To quantify such effect, one may define the nuclear modification factor $R_{AA}$ as follows:
\begin{eqnarray}
R_{AA}(p_T, y, \psi) = \frac{1}{N_{\rm coll}} \frac{dN_{AA}/dp_Tdyd\psi}{dN_{AA}/dp_Tdyd\psi} \,,
\end{eqnarray}
where $N_{\rm coll}$ is the average number of binary collisions for a given collision centrality bin.
Fig. \ref{fig_RAA_ID} shows the nuclear modification factors $R_{AA}$ for charged hadrons, direct photons and other particles; the left panel shows PHENIX measurements for most central Au-Au collisions at $200$~AGeV at RHIC, and the right shows CMS measurement for most central Pb-Pb collisions at $2.76$~ATeV at the LHC.
One can see that high $p_T$ hadron yields are strongly suppressed in A-A collisions compared to binary collision scaled p-p collisions.
In contrast, $R_{AA}$ for high $p_T$ photons is consistent with unity.
This means that the observed strong nuclear modification for high $p_T$ hadron production is a final state effect from the interaction of hard partonic jets with the hot and dense medium.

There are many other experimental signatures (observables) for jet quenching at RHIC and the LHC, such as the modification of the correlated back-to-back dihadrons and high $p_T$ photon-hadron pairs \cite{Adler:2002tq, Adams:2006yt, Adare:2009vd, Abelev:2009gu}, the suppression of reconstructed jet and dijet production \cite{Aad:2010bu,Chatrchyan:2011sx,Chatrchyan:2012gt,Aad:2014bxa}, and the nuclear effect on jet fragmentation and jet shape functions \cite{Chatrchyan:2012gw, Chatrchyan:2013kwa, Aad:2014wha}.
Similar to the single inclusive hadron $R_{AA}$, one may define the nuclear modification factors for these jet quenching obervables.

\subsection{General framework for jet quenching study}
\label{sec_general_framework}

In perturbative QCD, processes that involve large momentum transfer can be described as the convolution of the parton distribution functions (PDFs), hard partonic scattering process, and the final state fragmentation function (FFs).
For example, the cross section for the single inclusive hadrons at high $p_T$ in p-p collisions may be calculated as follows \cite{Jager:2002xm}:
\begin{eqnarray}
d\sigma_{pp \to hX} &&\!\! \approx \sum_{abj} \int dx_a \int dx_b \int dz_j f_{a/p}(x_a, \mu_f) \otimes f_{b/p}(x_b, \mu_f)
\nonumber\\ &&\!\! \otimes \ d\sigma_{ab\to jX}(\mu_f, \mu_F, \mu_R) \otimes D_{j\to h}(z_j, \mu_F) \,,
\end{eqnarray}
where $f_{a/p}(x_a, \mu_f)$ and $f_{b/p}(x_b, \mu_f)$ are two PDFs with $x_a$ and $x_b$ the momentum fractions of the incoming partons, $d\sigma_{ab\to jX}$ is the parton scattering cross section, and $D_{j\to h}(z_j, \mu_F)$ is the FF for the parton $j$ to the hadron $h$, with $z_j$ is the momentum fraction of the outgoing hadron.
There are three momentum scales involved here: the factorization scales $\mu_f$ and $\mu_F$ and the renormalization scale $\mu_R$; they are usually taken to be the same as a typical hard scale ($Q$) involved in the process, such as the hadron $p_T$.
The PDFs and FFs are non-perturbative and universal functions, and obey the Dokshitzer-Gribov-Lipatov-Altarelli-Parisi (DGLAP) equations for their scale evolutions \cite{Dokshitzer:1977sg, Gribov:1972ri, Altarelli:1977zs}.
They are usually determined by the global fit to $e^+e^-$ experiments, deep inelastic scatterings (DIS) and p-p collisions, etc.

\begin{figure}
\begin{center}
\includegraphics*[width=10.0cm, height=10.0cm]{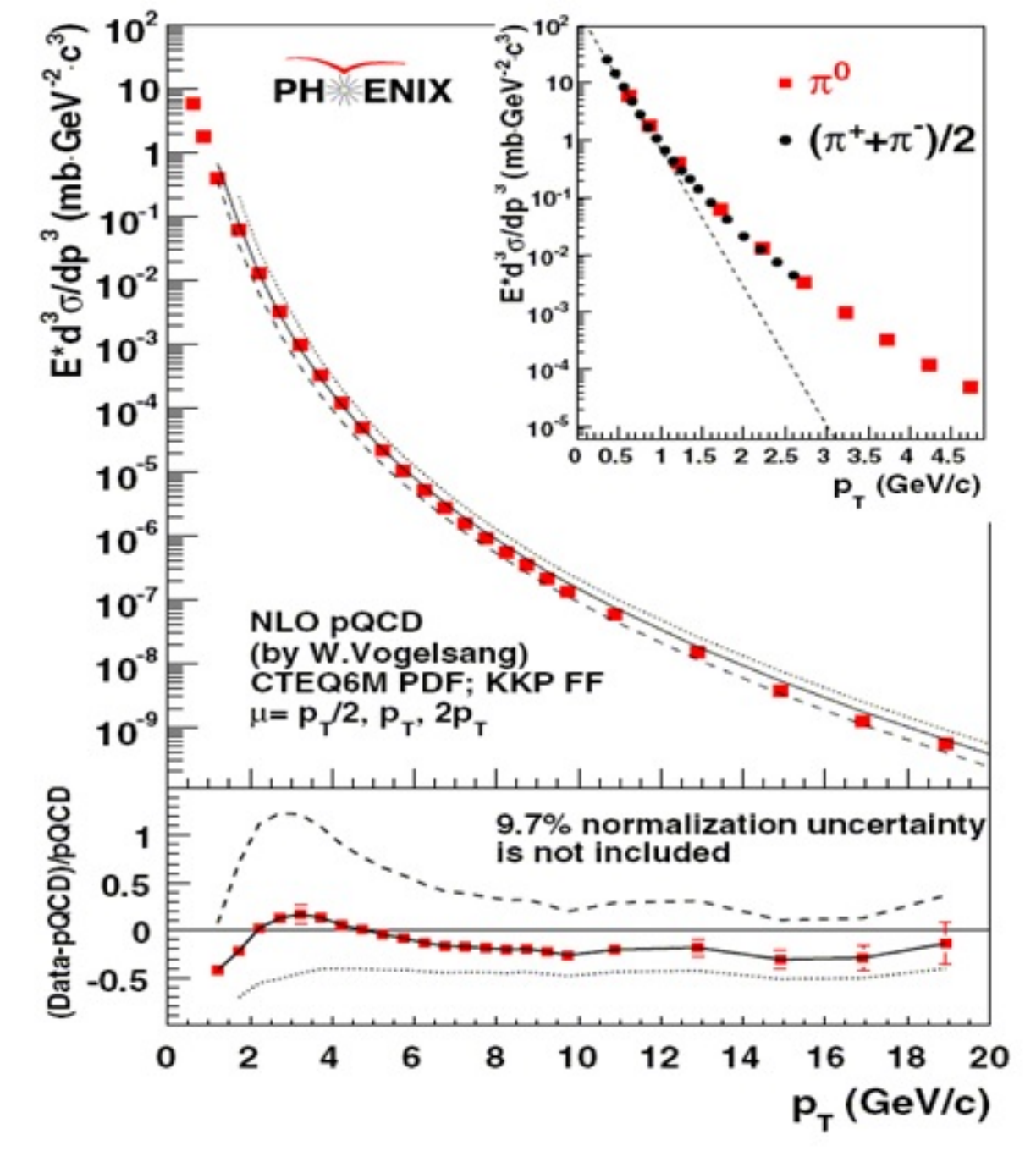}
\caption{(Color online) The cross section for inclusive hadrons  at high $p_T$ in p-p collisions at RHIC measured by PHENIX, compared to NLO perturbative QCD calculations (from Ref. \cite{Adare:2007dg}).
}
\label{fig_PHENIX_hadron_NLO}
\end{center}
\end{figure}

Fig. \ref{fig_PHENIX_hadron_NLO} shows the production cross section for single inclusive hadrons at high $p_T$ in elementary p-p collisions at RHIC energies.
One can see that the high $p_T$ hadron production can be described quite well by the next-to-leading order (NLO) pertubative QCD calculations \cite{Jager:2002xm}.
This indicates the properties of hard jets in vacuum are well understood from both experimental and theoretical sides, which serves as the baseline for studying jet-medium interaction and the nuclear modification of hard jets in relativistic heavy-ion collisions.

When studying jet quenching in relativistic heavy-ion collisions, two nuclear effects need to be taken into account.
One is the nuclear modification of PDFs, i.e., the PDF in nucleus $f_{a/A}$ is different from the free proton PDF $f_{a/p}$; this is the initial state nuclear effect, and often called cold nuclear matter (CNM) effect.
CNM effect can be included by defining the nuclear modification factor $R_i^A(x, Q^2)$ for the PDF: $R_a^A(x, Q^2)  = f_{a/A}(x, Q^2)/f_{a/p}(x, Q^2)$, which is often obtained by the global fit to DIS, p-A and d-A collisions.
Currently several parameterizations of PDF nuclear modification factors $R_i^A(x, Q^2)$ are available (e.g, EPS09 \cite{Eskola:2009uj}, HKN07 \cite{Hirai:2007sx}, and nDS \cite{deFlorian:2003qf}).

The second effect is due to the production of the hot and dense QGP medium in relativistic heavy-ion collisions, which we may call hot nuclear matter (HNM) effect.
Partonic jets produced from the initial hard scatterings have to travel through and interact with the produced QGP before fragmenting into final observed hadrons.
With the inclusion of both cold and hot nuclear effects, the single inclusive hadron production in heavy-ion collisions may be calculated as follows:
\begin{eqnarray}
d\tilde{\sigma}_{AB \to hX} \approx \sum_{abjj'} f_{a/A}(x_a) \otimes f_{b/B}(x_b) \otimes d\sigma_{ab\to jX} \otimes P_{j\to j'}(p_{j'}|p_j) \otimes D_{h/j'}(z_{j'})\,, \ \ \
\end{eqnarray}
Here $P_{j \to j'}(p_{j'}|p_j)$ describes the hot nuclear matter effect, i.e., the interaction of the hard partons $j$ with the colored medium.
It is noted that although the above factorized formula has been widely used in phenomenological studies of jet quenching in heavy-ion collisions, there is currently no formal proof of such factorization yet. The assumption of factorization is consistent with various jet quenching studies.

%\subsection{Parton energy loss formalisms}
\subsection{Radiative and collisional jet energy loss}

\begin{figure}
\begin{center}
\includegraphics*[width=5.9cm]{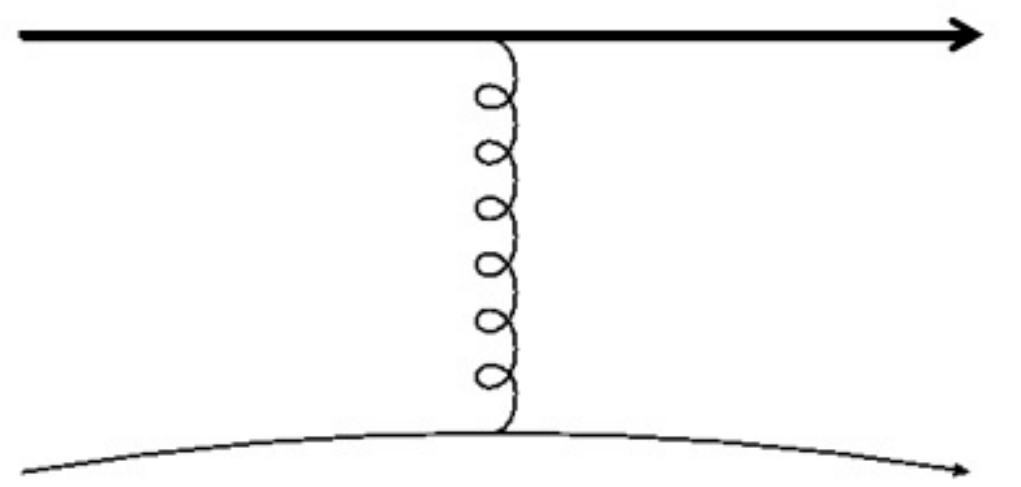}
\includegraphics*[width=6.6cm]{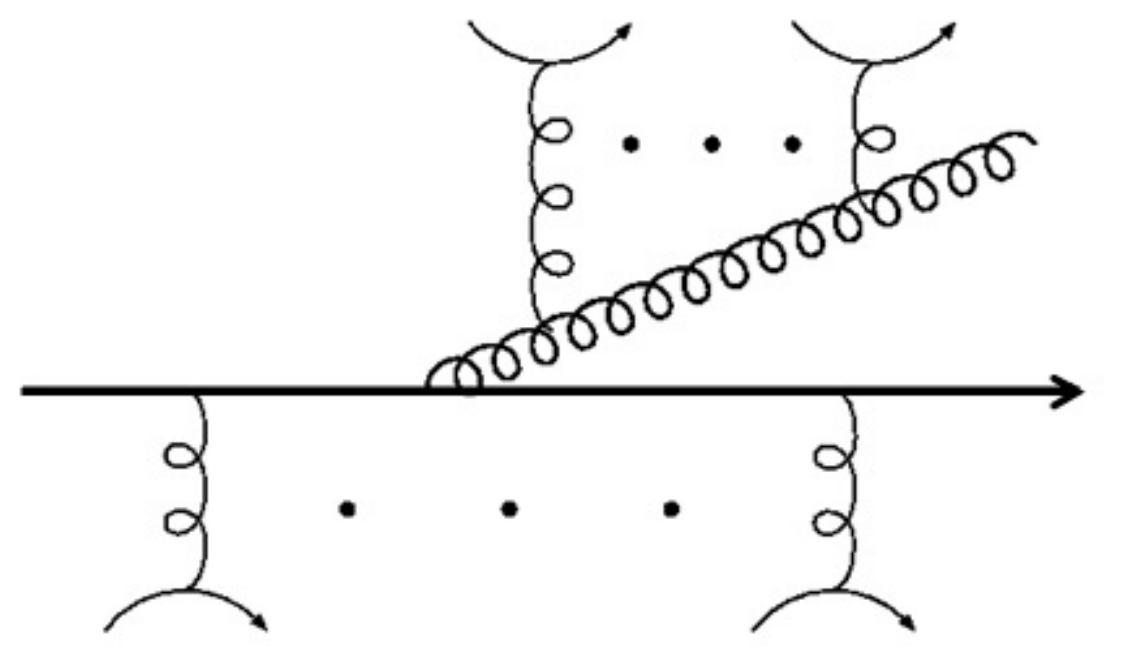}
\caption{(Color online) Typical diagrams for collisional (left panel) and radiative (right panel) energy losses of a hard parton propagating through a nuclear medium.
}
\label{fig_coll_rad}
\end{center}
\end{figure}

Hard partons may lose energy in hot and dense nuclear medium via a combination of elastic collisions with the medium constituents and inelastic radiative process.
Fig. \ref{fig_coll_rad} shows two typical diagrams for calculating parton energy loss originating from elastic (left panel) and radiative (right panel) processes.
The energy loss experienced by the primary parton in $2 \to 2$ elastic collisions with the medium constituents is usually called collisional or elastic energy loss.
The additional in-medium radiation induced by multiple scatterings may take away a fraction of energy from the primary parton, which is usually called radiative energy loss.

Radiative energy loss process has been regarded as the most important component in studying jet quenching in relativistic heavy-ion collisions.
Even in vacuum, hard partons produced from the early stage scatterings will undergo splitting processes.
In a hot and dense nuclear medium, the parton splitting processes will be modified due to the rescatterings of the propagating parton with the medium constituents.
A number of parton energy loss approaches have been developed to study medium-induced radiative process, namely Baier-Dokshitzer-Mueller-Peigne-Schiff-Zakharov (BDMPS-Z) \cite{Baier:1996kr,Baier:1996sk,Zakharov:1996fv}, Gyulassy, Levai and Vitev (GLV) \cite{Gyulassy:1999zd,Gyulassy:2000fs,Djordjevic:2008iz}, Amesto-Salgado-Wiedemann (ASW) \cite{Wiedemann:2000za,Wiedemann:2000tf}, higher twist (HT) \cite{Guo:2000nz,Wang:2001ifa,Majumder:2009ge} and Arnold-Moore-Yaffe (AMY) \cite{Arnold:2001ba,Arnold:2002ja,CaronHuot:2010bp} formalisms.
One may refer to Ref. \cite{Armesto:2011ht} for a detailed comparison of different parton energy loss formalisms.

In the study of radiative parton energy loss, one key quantity is the single emission kernel (e.g., the differential gluon radiation spectrum $dN_g/d\omega dk_\perp^2 dt$).
Various approaches make different assumptions about the traversed medium and utilize different methods for treating multiple scatterings.
One important effect in medium-induced emission or splitting processes is the Landau-Pomeranchuk-Migidal (LPM) effect \cite{Landau:1953um,Migdal:1956tc}.
For collinear or small angle radiation, a finite period of time is required to complete the radiation process; this time is called the formation time $\tau_f \sim 2\omega/k_\perp^2$, with $\omega$ and $k_\perp$ the energy and transverse momentum of the radiation.
If the formation time is larger than the mean free path $\lambda$ of the propagating parton, the multiple scatterings on the propagating parton can no longer be treated as independent.
Such quantum interference between successive scatterings is called the LPM effect which will lead to the suppression of the radiation spectrum compared to the Bethe-Heitler incoherent multiple scattering limit.
For the QCD case, since the radiated gluons carry color charge, the medium modification of the radiation spectrum is more dominated by the rescatterings on the emitted gluons.

To obtain the multiple gluon emission, one common practice is the repeated application of single gluon emission kernel.
Such recipe neglects the interference between different emissions which should be included in a full calculation of multiple parton final state.
One popular method is the Poisson ansatz which has been widely used in GLV and ASW calculations \cite{Salgado:2003gb, Renk:2006sx, Wicks:2005gt}.
The key quantity is the probability distribution $P(\Delta E)$ of parton energy loss, which may be obtained as follows:
\begin{eqnarray}
P(\Delta E) = \sum_{n=0}^{\infty} \frac{e^{-\langle N_g \rangle}}{n!} \left[ \prod_{i=1}^{n} \int d\omega \frac{dN_g(\omega)}{d\omega} \right] \delta\left( \Delta E - \sum_{i=1}^{n} \omega_i \right)\,,
\end{eqnarray}
where $dN/d\omega$ is the single gluon emission spectrum and $\langle N_g \rangle = \int d\omega dN_g/d\omega$ is the mean number of radiated gluons.
In the AMY formalism, the following rate equations are solved for the parton momentum distributions $f(p)=dN(p)/dp$  \cite{Jeon:2003gi, Qin:2007zzf, Qin:2007rn}:
\begin{eqnarray}
\frac{df(p,t)}{dt} = \int dk \left[ f(p+k,t) \frac{d\Gamma(p+k,k,t)}{dkdt} - f(p,t) \frac{d\Gamma(p,k,t)}{dkdt} \right],
\end{eqnarray}
where $d\Gamma(p,k,t)/dkdt$ is the transition rate for a parton with momentum $p$ to lose momentum $k$.
In the HT formalism, one solves the following DGLAP-like evolution equations for the medium-modified fragmentation function $\tilde{D}(z,Q^2)$ \cite{Majumder:2009zu, Qin:2009gw, Majumder:2011uk, Chang:2014fba}:
\begin{eqnarray}
\frac{\partial \tilde{D}(z, Q^2, q^-)}{\partial\ln Q^2} = \frac{\alpha_s}{2\pi} \int \frac{dy}{y} P(y) \int d\zeta^- K(\zeta^-, Q^2, q^-, y) \tilde{D}(z/y, Q^2, q^-y)\,,
\end{eqnarray}
where $q^-$ is the jet light-cone energy, $y$ the energy fraction of the radiation, and $\zeta^-$ is the jet location.
$P(y)$ is the vacuum splitting function and $K(\zeta^-, Q^2, q^-, y)$ is the parton-medium interaction kernel.

\begin{figure}
\begin{center}
\includegraphics*[width=6.25cm]{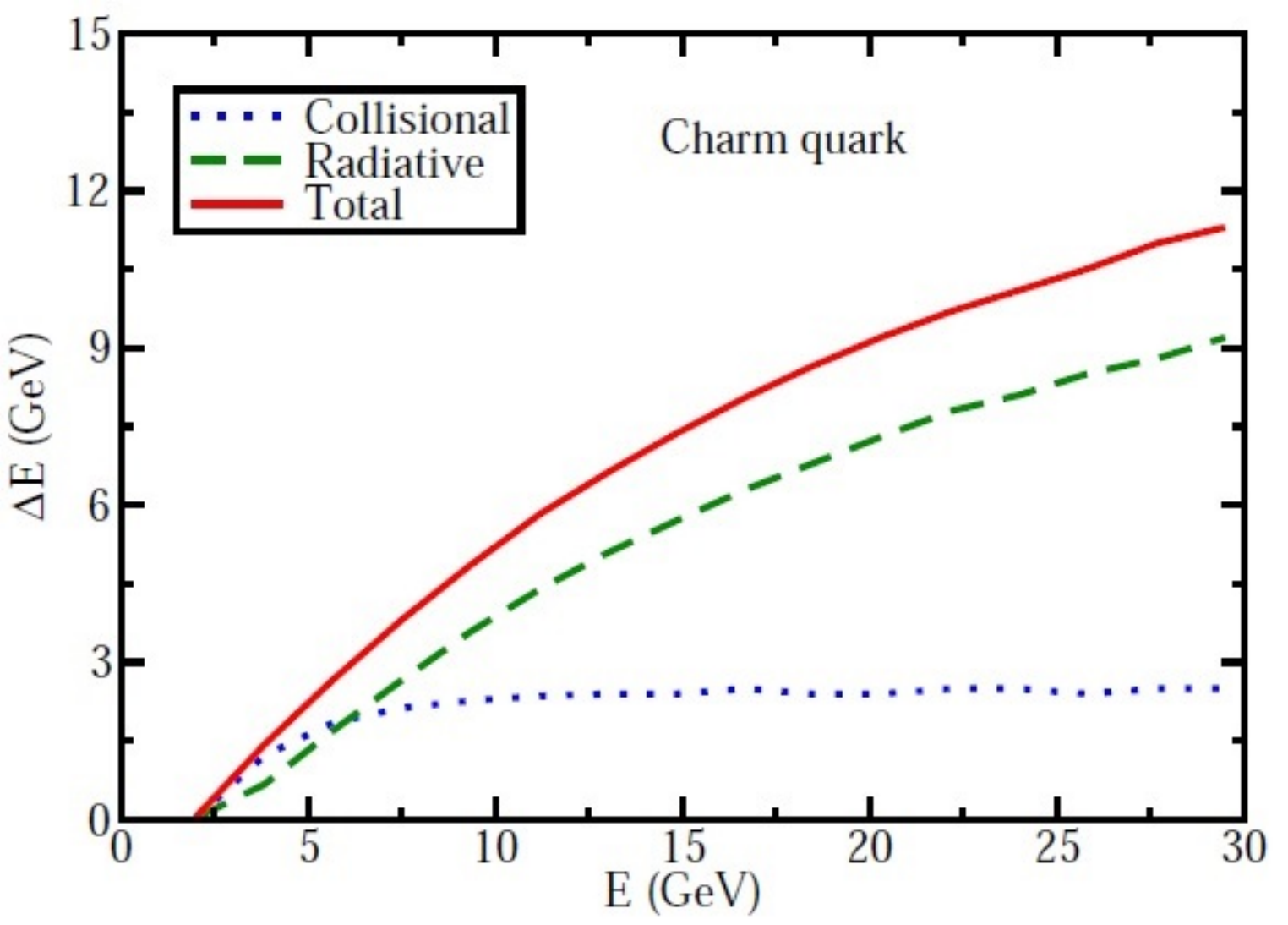}
\includegraphics*[width=6.25cm]{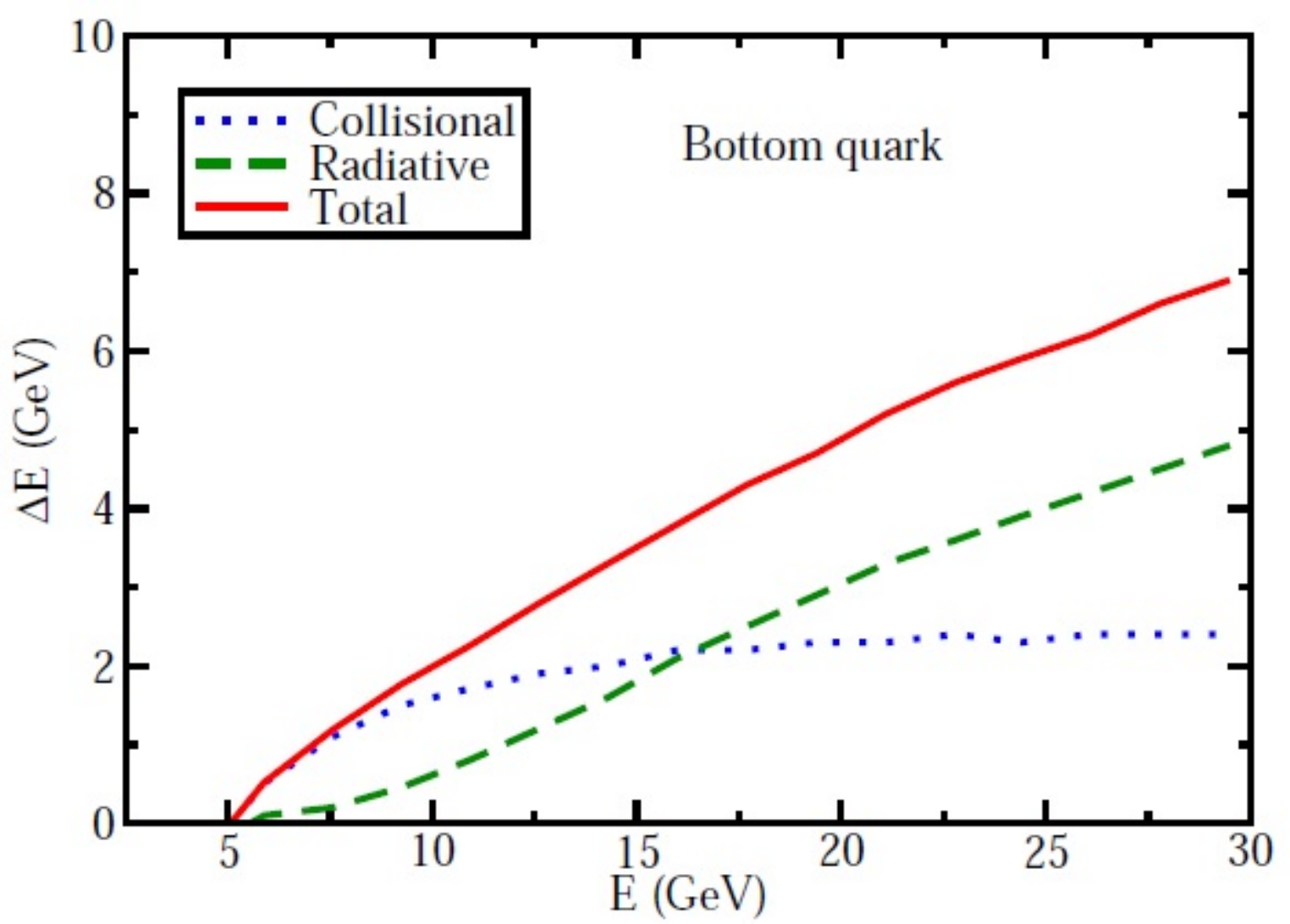}
\caption{(Color online)
Right: Average energy loss of charm (left panel) and bottom (right panel) quarks passing through the nuclear medium created in most 0-7.5\% central Pb-Pb collisions at $2.76$~ATeV at the LHC (from Ref. \cite{Cao:2013ita}).
}
\label{fig_Qin_rad_vs_coll}
\end{center}
\end{figure}

Collisional energy loss of hard partons propagating through the nuclear matter was first studied by Bjorken \cite{Bjorken:1982tu}.
Compared to medium-induced radiation, collisional energy loss is usually considered to be small for light flavor (leading) partons, especially at high energies  \cite{Wicks:2005gt,Qin:2007rn,Schenke:2009ik}.
However, collisional energy loss may give sizable contribution to nuclear modification factor $R_{AA}$ at RHIC and the LHC energies \cite{Wicks:2005gt,Qin:2007rn}.
In contrast, elastic collisions are usually considered as the dominant mechanism for heavy quark energy loss, especially at low and intermediate energy regimes \cite{Moore:2004tg, Mustafa:2004dr}.
This is because the phase space for collinear medium-induced radiation is reduced by the presence of the finite masses of heavy quarks (usually called the dead-cone effect \cite{Dokshitzer:2001zm}).
This mass effect will diminish when going to high energy regimes where heavy quarks become ultra-relativistic as well and behave more like light flavor partons \cite{Cao:2012au, Uphoff:2012gb, Abir:2012pu, Nahrgang:2014vza}.
This can be seen in Fig. \ref{fig_Qin_rad_vs_coll} which compares the radiative and collisional contributions to the energy loss of charm (left panel) and bottom (right panel) quarks.
We note that collisional energy loss may play an essential role in the study of the nuclear modification of full jets \cite{Qin:2010mn} and the medium response to the propagation of hard jets \cite{Qin:2009uh,Neufeld:2009ep}; these will be discussed in later subsections.

\subsection{Jet quenching phenomenology at RHIC and the LHC and quantitative extraction of $\hat{q}$}

In recent years, various phenomenological studies on jet quenching have been performed for a wealth of experimental observables.
Much effort has been made to the quantitative extraction of various jet transport parameters \cite{Bass:2008rv}, such as the transverse momentum diffusion rate $\hat{q} = d\langle \Delta p_\perp^2 \rangle/dt$ \cite{Baier:1996kr}, the elastic energy loss rate $\hat{e}=dE/dt \approx dp_\parallel/dt$, etc \cite{Majumder:2007hx, Qin:2012fua}.
At leading order (LO), these transport coefficients quantify the transverse and longitudinal momentum transfers experienced by the propagating partons via $2\to 2$ elastic collisions with the medium constituents.

Recently a significant collaborative effort was carried out within the framework of JET Collaboration; systematic phenomenological studies were performed on the experimental data on the nuclear modification of single inclusive hadrons at large $p_T$ in heavy-ion collisions at the RHIC and the LHC \cite{Burke:2013yra}.
Five different existing approaches to parton propagation and medium-induced energy loss in dense medium were used for this systematic survey: McGill-AMY \cite{Qin:2007rn, Qin:2009bk}, Martini-AMY \cite{Young:2011ug}, HT-M \cite{Majumder:2011uk}, HT-BW \cite{Chen:2011vt}, DGLV-CUJET \cite{Xu:2014ica}.
The space-time profile of the QGP medium was simulated by the (2+1)-dimensional or (3+1)-dimensional hydrodynamic models.
The goal of this systematic study is to use the constraint from the experimental data and to quantitatively extract the jet quenching parameter $\hat{q}$ and its systematic uncertainties (including model dependence).

\begin{figure}
\begin{center}
\includegraphics*[width=10.0cm, height=7.0cm]{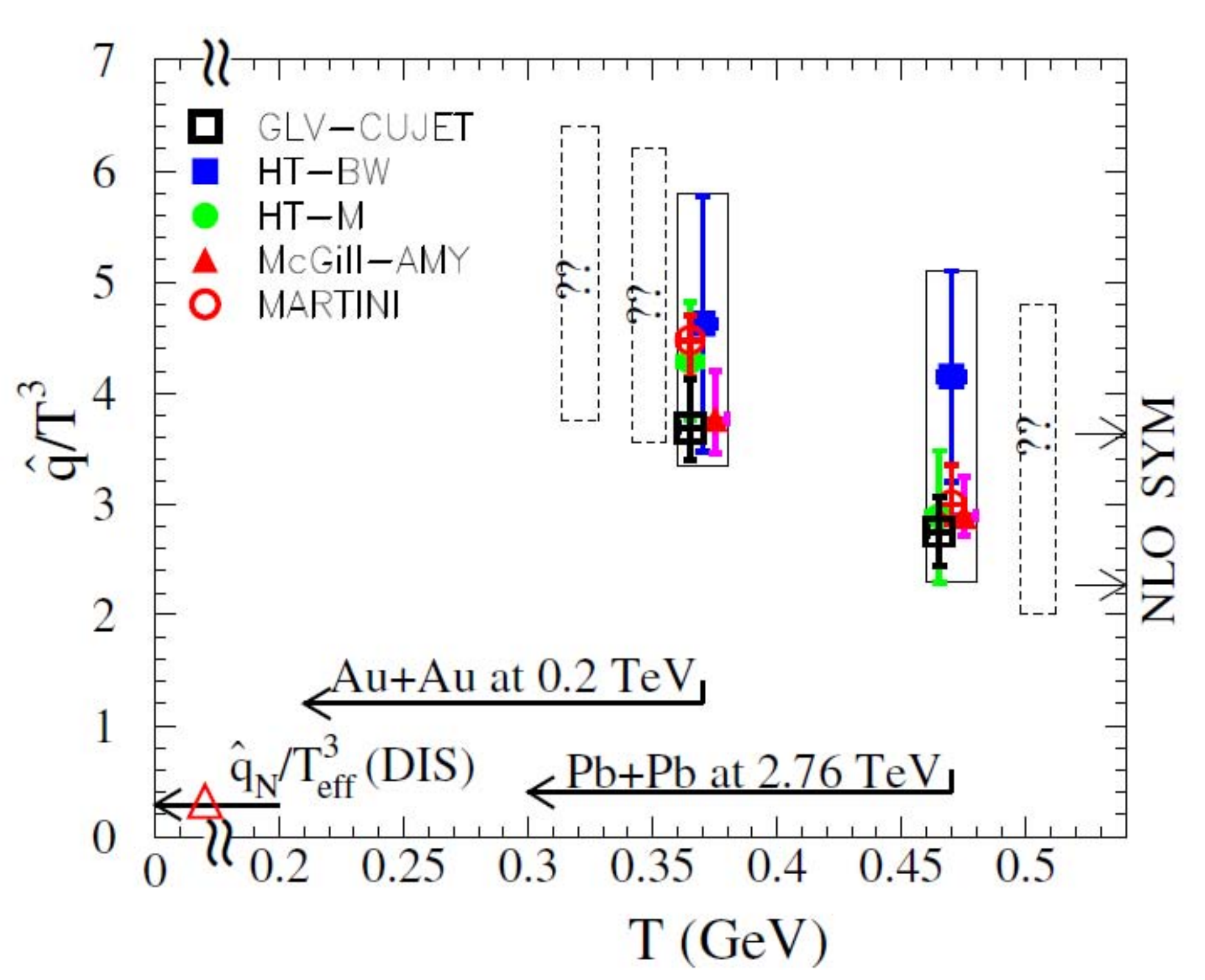}
\caption{(Color online) The values of scaled jet transport parameter $\hat{q}/T^3$ extracted by JET Collaboration by using single inclusive hadron suppression factor $R_{AA}$ at both RHIC and LHC.
The values are for a $10$~GeV quark jet at the center of most central collisions at an initial proper time $\tau_0 = 0.6$~fm/c.
The figure is taken from Ref. \cite{Burke:2013yra}.
}
\label{fig_JET_qhat}
\end{center}
\end{figure}

By comparing each model calculation with the experimental data and fixing the model parameters, the effective jet transport coefficient $\hat{q}$ may be obtained.
Fig. \ref{fig_JET_qhat} shows the extracted values of $\hat{q}$ scaled by $T^3$ for a quark jet with energy $10$~GeV at the highest temperatures reached in the most central Au-Au collisions $200$~AGeV at RHIC and Pb-Pb collisions $2.76$~ATeV at the LHC.
Using the measured single hadron $R_{AA}$, the range of the values for $\hat{q}/T^3$ is obtained as:
\begin{eqnarray}
\frac{\hat{q}}{C_s T^3} = \left\{
                                \begin{array}{ll}
                                  3.5 \pm 0.9\,, & \hbox{$T \approx 370$ {\rm MeV (at RHIC)},} \\
                                  2.8 \pm 1.1\,, & \hbox{$T \approx 470$ {\rm MeV (at the LHC)}.}
                                \end{array}
                              \right.
\end{eqnarray}
This gives $\hat{q} \approx$ 1.2~GeV$^2/$fm at RHIC and $\hat{q} \approx$ 1.9~GeV$^2/$~fm at the LHC.

One can clearly see the strong temperature dependence for the scaled parameter $\hat{q}/T^3$, which may be explored by extending the current study to future higher energy Pb-Pb collisions at the LHC and lower energy collisions at RHIC.
The expected values of $\hat{q}/T^3$ at 0.063~ATeV, 0.130~ATeV and 5.5~ATeV are shown in the figure.
Also shown is the value of $\hat{q}_N/T_{\rm eff}^3$ in cold nuclei, extracted from jet quenching studies in DIS \cite{Majumder:2004pt}.
One can see that the values of jet parameter $\hat{q}$ in hot QGP medium are much higher than those in cold nuclei.
The values of $\hat{q}/T^3$ from a NLO AdS/CFT calculation is also shown for comparison \cite{Zhang:2012jd}, and one can see that they are within the range of $\hat{q}$ values from JET Collaboration.
Note that the SYM values quoted here have been obtained for $\alpha_{SYM} =$~0.22-0.31 and included the effect due to different numbers of degrees of freedom in SYM and QCD.

In the future, one may include more theoretical model calculations and utilize more jet quenching observables in such systematic phenomenological studies \cite{Renk:2011aa}.
Besides $\hat{q}$, other jet transport coefficients (such as $\hat{e}$) may also play significant roles in jet-medium interaction and medium-induced parton energy loss \cite{Majumder:2007hx, Qin:2012fua, Qin:2014mya, ColemanSmith:2012xb}.
To estimate the systematic uncertainties in the LO calculations, it is essential to develop a fully NLO framework for studying jet propagation and modification in dense nuclear matter.
Recently the NLO radiative correction to transverse momentum broadening and the renormalization of jet quenching parameter $\hat{q}$ have been investigated in Ref. \cite{Wu:2011kc, Liou:2013qya, Iancu:2014kga, Blaizot:2014bha, Kang:2013raa, Kang:2014ela}.
All these important ingredients should be included in the future study of jet quenching in heavy-ion collisions.

\subsection{Jet quenching from Lattice QCD}

While lattice QCD cannot provide a full description of the dynamical evolution of heavy-ion collisions, it can provide important guidance for phenomenological jet quenching studies, e.g., many transport coefficients (such as $\hat{q}$) may in principle be computed using lattice QCD.
Ref. \cite{Majumder:2012sh} carried out the first lattice QCD calculation of jet quenching parameter $\hat{q}$ using the operator product expansion method; the physical $\hat{q}$ is related to an infinite series local operators in an unphyical regime of jet momenta via dispersion relations.
The calculation was performed in quenched $SU(2)$ and the extension to $SU(3)$ with $2$ flavors of quarks produced $\hat{q}$=1.3-3.3~GeV$^2$/fm for a gluon jet at a temperature $T=400$~MeV, which is similar to the values from JET Collaboration.
Recently within a dimensionally reduced effective theory (electrostatic QCD), Ref. \cite{Panero:2013pla} obtained a value of $\hat{q}$=6~GeV$^2$/fm at RHIC energies, which is about twice the value compared to NLO perturbative QCD calculation \cite{CaronHuot:2008ni}.

\begin{figure}
\begin{center}
\includegraphics*[width=12.0cm]{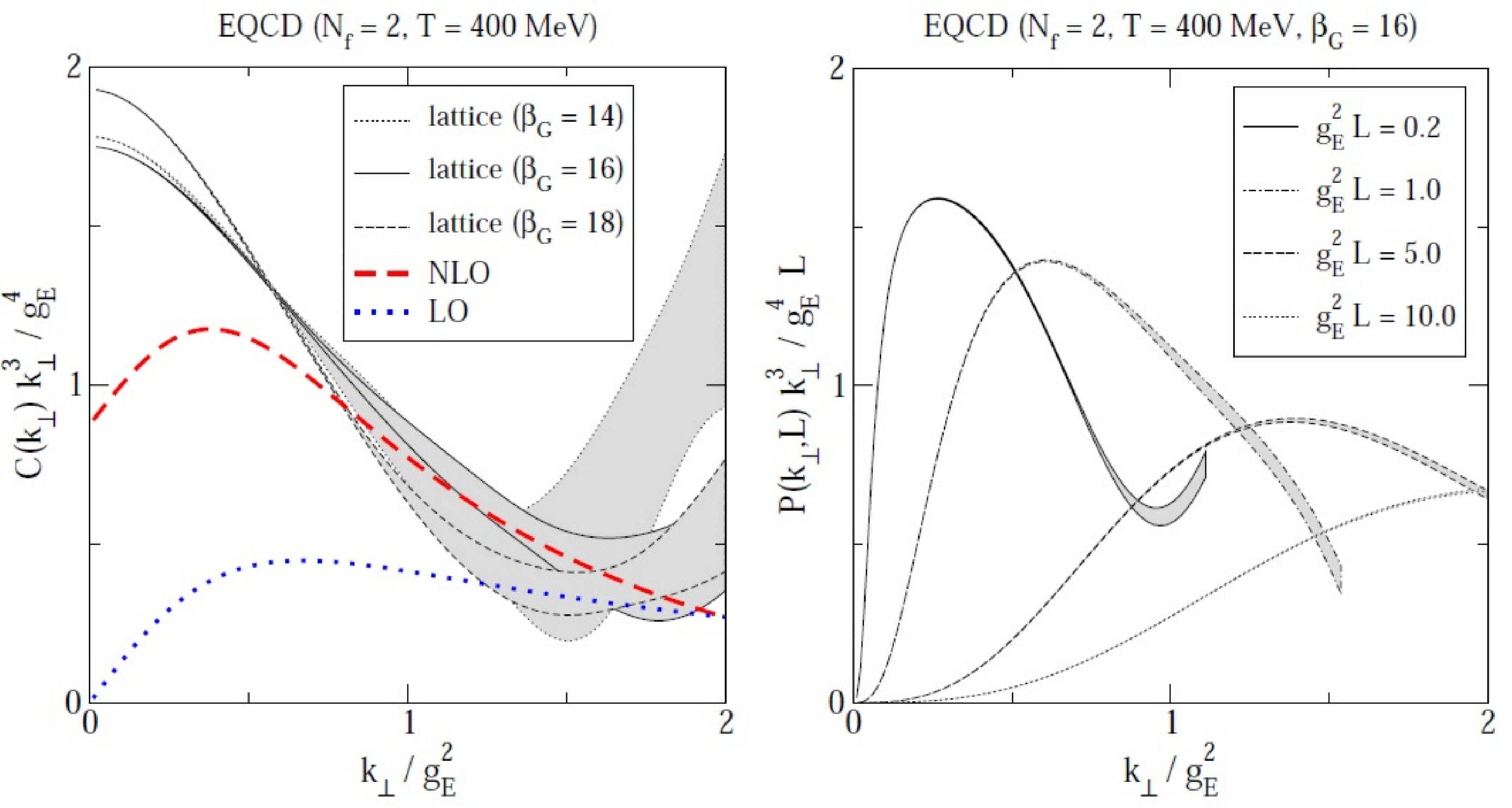}
\caption{(Color online) The collision kernel $C(k_\perp)$ (left panel) and the probability distribution $P(k_\perp,L)$ (right panel) compared with the LO and NLO calculation results from Ref. \cite{CaronHuot:2008ni}.
$g_E^2 \sim g^2T$ is the effective coupling of the dimensionally reduced effective theory and $\beta_G$ is a dimensionless number related to the lattice spacing: $\beta_G = 2N_c/(g^2T a)$.
The figures are taken from Ref. \cite{Laine:2013apa}.
}
\label{fig_EQCD_Ck_Pk}
\end{center}
\end{figure}

We note that the jet quenching parameter $\hat{q}$ is the second moment of the probability distribution $P(k_\perp, L)$ for the transverse momentum transfer,
\begin{eqnarray}
\hat{q} = \frac{1}{L} \int \frac{d^2k_\perp}{(2\pi)^2} k_\perp^2 P(\mathbf{k}_\perp, L) \,.
\end{eqnarray}
More detailed information about jet-medium interaction are contained in the full distribution of the probability distribution $P(k_\perp)$ or the transverse collision kernel $C(k_\perp)$.
In Ref. \cite{Laine:2013apa}, a first-principle calculation of the collision kernel $C(k_\perp)$ was carried out (see Fig. \ref{fig_EQCD_Ck_Pk}).
At small transverse momentum $k_\perp$, the shape of $C(k_\perp)$ was found to be consistent with Gaussian distribution.
Since the calculation only includes the soft modes, it breaks down when $k_\perp$ becomes large.

The Gaussian form for $P(k_\perp)$ and $C(k_\perp)$ at small $k_\perp$ is expected from both weakly-coupled and strongly-coupled calculations.
However the large $k_\perp$ behaviors are very different in two scenarios: for weakly coupled QGP, the power tail for large $k_\perp$ is proportional to $1/k_\perp^2$ , whereas in $\mathcal{N} = 4$ SYM theory (in the large-$N_c$ and strong coupling limits), there is no power tail at all \cite{D'Eramo:2012jh}.
One interesting task is whether the full distribution of the collision kernel can be obtained from phenomenological jet quenching studies (as well as from lattice QCD calculation).

\subsection{Full jet evolution and energy loss}

The basic idea of full jet reconstruction is to recombine final state jet fragments and obtain the information about the original hard partons and investigate the medium modification effects.
Full jets are expected to provide more differential information about jet-medium interaction than leading hadron observables since they include both leading and sub-leading jet fragments.
Early full jet studies in heavy-ion collisions were performed in Au-Au and Cu-Cu collisions at RHIC by both STAR and PHENIX Collaborations \cite{Ploskon:2009zd, Lai:2009zq}.
In spite of large experimental uncertainties and strong dependence on jet reconstruction algorithms, substantial nuclear modification of full jets in heavy-ion collisions have been observed as compared to p-p collisions.
The strong dependence on the jet resolution parameter $R$ for the nuclear modification of jet production indicated the broadening of full jets due to the interaction between jets and the hot and dense nuclear medium \cite{Vitev:2009rd}.

\begin{figure}
\begin{center}
\includegraphics*[width=6.5cm, height=4.6cm]{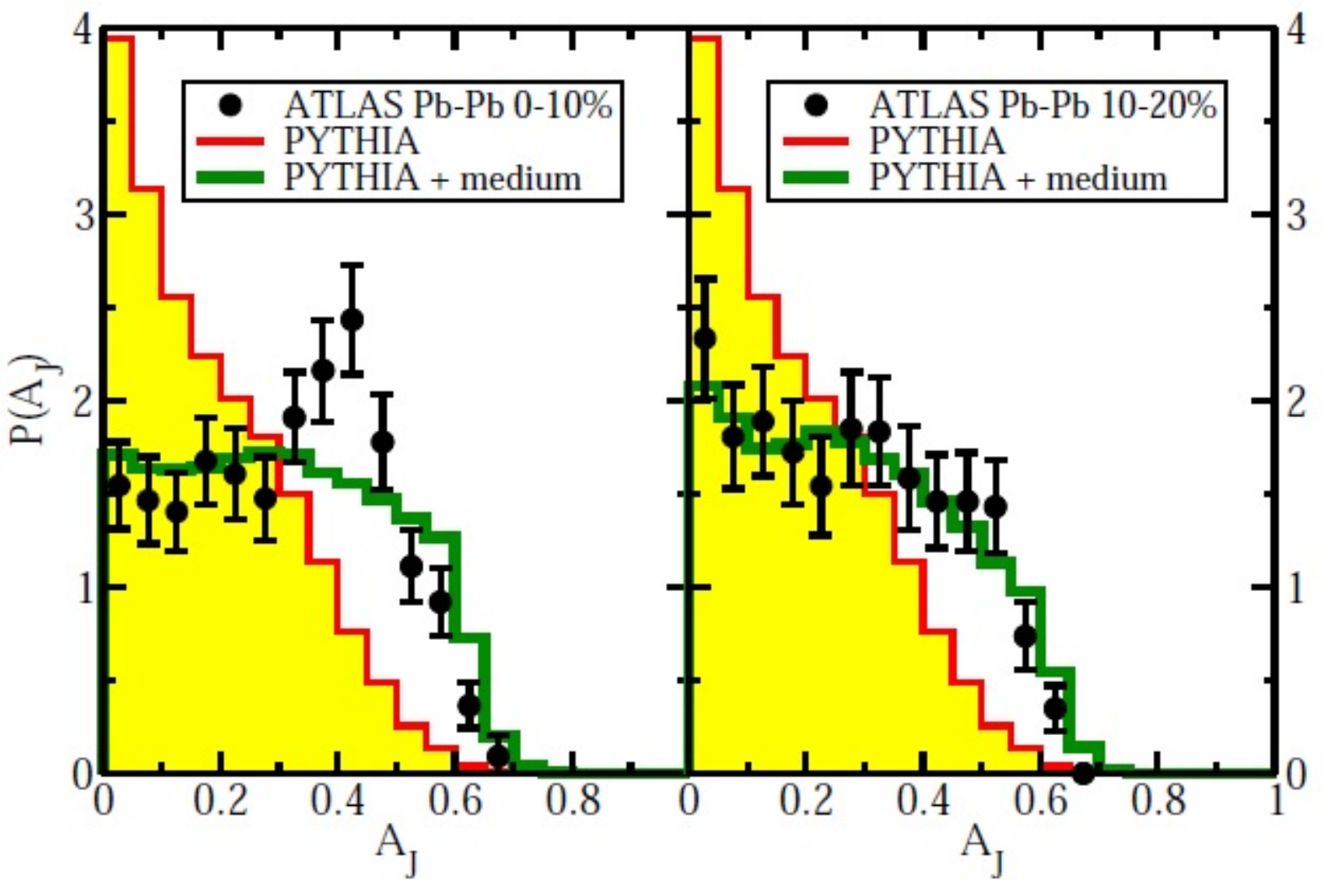}
\includegraphics*[width=6.0cm, height=4.6cm]{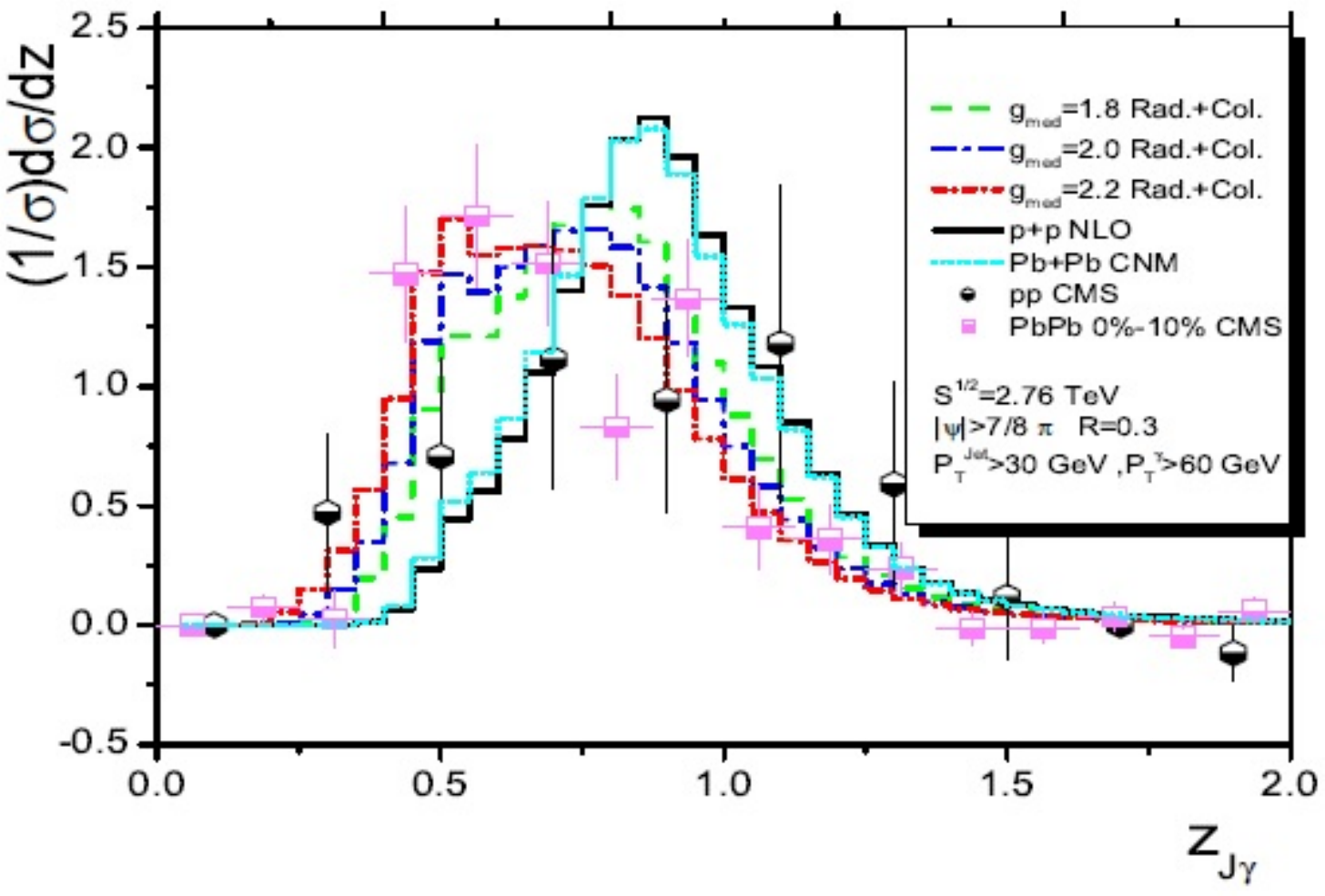}
\caption{(Color online) Left: The energy asymmetry $A_J = ({E_{T, 1} - E_{T, 2}})/({E_{T, 1} + E_{T, 2}})$ distribution for dijets in p-p and Pb-Pb collisions at $2760$~AGeV at the LHC (from Ref. \cite{Qin:2010mn}). Right: The energy asymmetry $z_{J\gamma} = E_{T, J} / E_{T, \gamma}$ distribution  for correlated isolated photons and jets in p-p and Pb-Pb collisions at $2760$~AGeV at the LHC (from Ref. \cite{Dai:2012am}).
}
\label{fig_QM_AJ}
\end{center}
\end{figure}

The launch of the LHC has increased the center of mass energy by more than a factor of 10 compared to the top collision energies at RHIC, which enables us to investigate the medium modification on the propagation of jets with transverse energies over a hundred GeV.
The first measurements of the nuclear modification of full jets at the LHC is the correlated back-to-back jet pairs.
We observed a strong centrality dependence for the modification on dijet transverse energy imbalance $A_J = ({E_{T, 1} - E_{T, 2}})/({E_{T, 1} + E_{T, 2}})$ distribution in Pb-Pb collisions as compared to p-p collisions, while the distribution of their relative azimuthal angles is largely unmodified.
Similar results have been obtained for the full jets correlated with high $p_T$ direct photons in Pb-Pb collisions at the LHC \cite{Chatrchyan:2012gt}.
These results indicate that the subleading jets may experience a significant amount of energy loss after passing through the produced hot and dense QCD matter.
Various jet energy loss calculations have been performed to explain the nuclear modification of the transverse energy imbalance between correlated dijets and photon-jet pairs \cite{Qin:2010mn,Young:2011qx,He:2011pd,Renk:2012cx,Zapp:2012ak,Dai:2012am,Qin:2012gp,Wang:2013cia} (e.g. see Fig. \ref{fig_QM_AJ}).

\begin{figure}
\begin{center}
\includegraphics*[width=12.0cm,height=4.5cm]{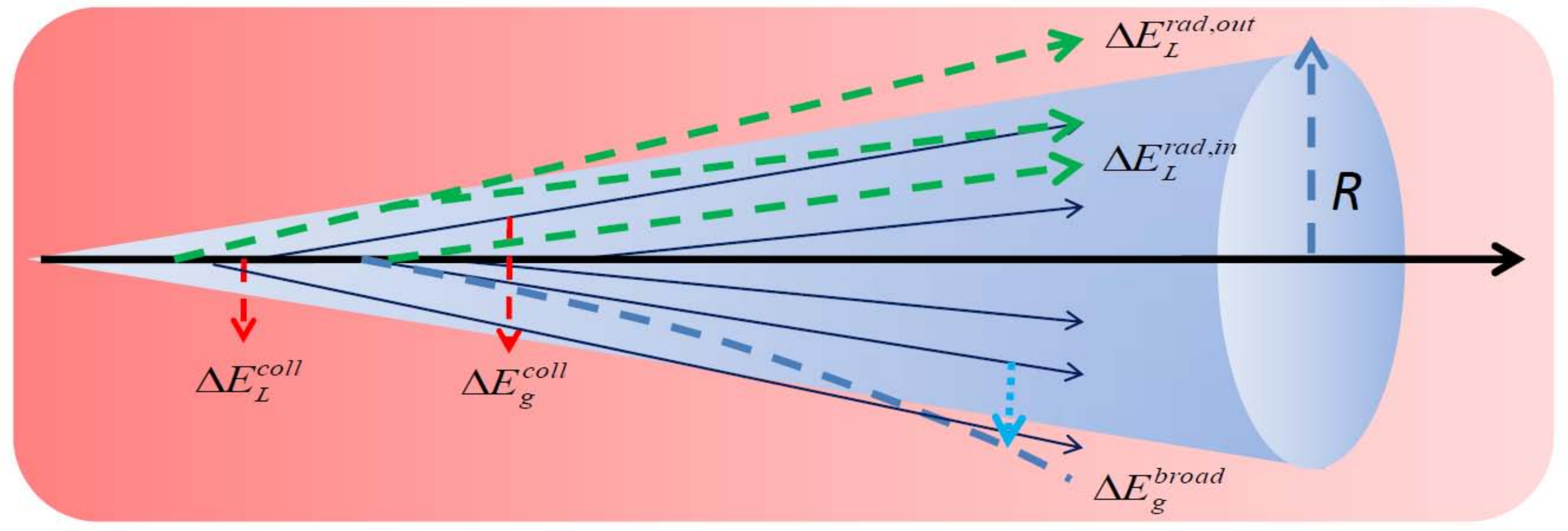}
\caption{(Color online) Illustration of the evolution of a full jet in a quark-gluon plasma. Various medium-induced processes that contribute to full jet energy loss and medium modification are shown.
}
\label{fig_full_jet_evolution}
\end{center}
\end{figure}

Fig. \ref{fig_full_jet_evolution} shows a schematic illustration of the evolution of a full jet in a QGP, where the thick solid arrowed line through the center represents the leading parton, and other lines represent the accompanying radiated gluons.
Compared to leading hardon observables, we need to consider a few additional ingredients when studying full jet evolution and energy loss.
Besides the primary parton, the radiated gluons may interact with the medium and lose energy in the process.
The radiated gluons may get deflected by the medium constituents as well; some of the radiations may be scattered out of the jet cone.
Therefore, the total energy loss of the full jet is the sum of elastic energy loss experienced by the leading parton and the radiated gluons, together with the gluons that are kicked out of the jet cone.

In the spirit of the above picture, Ref. \cite{Qin:2010mn} performed the first quantitative analysis of physical processes that are responsible for full jet energy loss. It solves the following transport equation for the radiated gluons:
\begin{eqnarray}
\frac{df_g(\omega, k_\perp, t)}{dt} = -\hat{e}\frac{\partial f_g}{\partial \omega} + \frac{1}{4} \hat{q} \nabla_{k_\perp}^2 f_g + \frac{dN_g^{\rm rad}}{d\omega dk_\perp^2 dt}\,,
\end{eqnarray}
where $f_g(\omega, k_\perp, t)$ is the three-dimensional momentum distribution of the accompanying gluons.
The first and second terms in the above equation describe the evolution of the radiative gluons which may transfer energy into the medium via elastic collisions and accumulate transverse momentum in the process.
The last term is a source term which represents the contribution from the medium-induced gluon radiation.
After they are produced, the medium-induced radiative gluons will interact with the medium, lose energy and accumulate transverse momentum during their propagation.

By solving the above equation, one may obtain the information about the full jets after passing through the medium and calculate the energy loss from the jet cone.
One may decompose the energy of the original full jet into a few parts:
\begin{eqnarray}
E_{\rm jet} = E_{\rm in} + E_{\rm lost} = E_{\rm in, rad} + E_{\rm out, rad} + E_{\rm out, brd} + E_{\rm th, coll}\,.
\end{eqnarray}
It is interesting that the most significant contribution is found to be the collisional energy loss experienced by the radiated gluons $E_{\rm th, coll}$ \cite{Qin:2010mn}.
The transverse momentum broadening of the radiative gluons $E_{\rm out, brd}$ also gives a sizable contribution to full jet energy loss.
Since the radiation is mainly dominated by small angle radiation, the contribution from the radiation directly outside the jet cone $E_{\rm out, rad}$ is small.
This finding is not surprising since one expects the soft components of the jet or the accompanying gluons at large angles to experience stronger medium modification than the inner hard core of the jet.
Such effect is also referred to as jet collimation \cite{CasalderreySolana:2010eh}.

Similar picture for full jet energy loss has been obtained in Ref. \cite{CasalderreySolana:2012ef,Blaizot:2013hx}.
It is argued that if the medium color field $\lambda$ varies over the jet transverse size $r_{\perp}^{\rm jet}$, the color coherence of the shower partons may be destroyed by the interaction with the medium.
Such color decoherence effect may greatly increase the phase space for soft and large angle radiation, as compared to the traditional BDMPS-Z radiative energy loss formalism.
Treating multiple gluon emissions as a probabilistic branching process, one may solve the following rate equation for the gluon momentum distribution $D(x, \tau) = x dN_g/dx$,
\begin{eqnarray}
\frac{\partial D(x, \tau)}{\partial \tau} = \int dz \kappa(z) \left[ \sqrt{\frac{z}{z}} D(\frac{x}{z}, \tau) - \frac{z}{\sqrt{x}} D(x, \tau) \right] \,.
\end{eqnarray}
Two terms in the above equation are gain and loss terms. Focusing on the small-$x$ behavior, one may take $\kappa(z) = 1$.
Given the initial condition $D(x, \tau=0) = \delta(x - 1)$,  one may obtain the solution for the gluon distribution: $D(x, \tau) \approx \frac{\tau}{\sqrt{x}} e^{-\pi \tau^2}$. The total energy contained in the spectrum may be calculated as, $\epsilon(\tau) = \int_0^1 dx D(x, \tau) \approx e^{- \pi \tau^2}$, and one can see that it is not conserved.
Therefore with increasing time, a substantial fraction of jet energy could be lost from the gluon spectrum.

To include the contribution from radiation at large angles, one may introduce a scale $x_0$, below which the radiation is directly outside jet cone.
One may further introduce a scale $x_{\rm th} = T/E$ to include the contribution that when the gluon energies are as low as as the medium energy scale, the radiated gluons may ``thermalize" and disappear from the jet.
So the total phase space for the branching gluons (and the total energy of the original jet) may be divided into three parts \cite{Iancu:2013ura}:
\begin{eqnarray}
E_{\rm jet} = E_{\rm in} + E_{\rm lost} = E_{\rm in}(x>x_0) + E_{\rm out}(x_{\rm th}<x<x_0) + E_{\rm flow}(x<x_{\rm th}).
\end{eqnarray}
The branching gluons with $x>x_0$ are inside the jet cone.
For $x_{\rm th} < x < x_0$, the radiations are directly outside the jet cone.
For $x<x_{\rm th}$, the gluons will thermalize and flow into the medium.
The largest contribution to full jet energy loss is found to be the energy flowing out of the gluon spectrum.
The radiation directly outside jet cone gives small contribution.

\subsection{Full jet substructure}

To fully characterize the jet quenching effects, we need to study not only the total energy of the full jets, but also the modification of the jet substructures.
Jet shape function $\rho(r)$ describes the radial distribution of the momentum carried by the jet fragments. It is defined as follows:
\begin{eqnarray}
\rho(r) = \frac{1}{\delta r} \frac{1}{N_{\rm jet}} \sum_{\rm jet}  \sum_{h} \frac{ p_T^h}{p_T^{\rm jet}} \theta\left[r_h - \left(r-\frac{\delta r}{2}\right)\right] \theta\left[ \left(r+\frac{\delta r}{2}\right) - r_h \right] \,,
\end{eqnarray}
where $r_h = \sqrt{(\eta_h - \eta_{\rm jet})^2 + (\phi_h - \phi_{\rm jet})^2}$, and $\delta r$ is the bin size.
The differential jet shape function is normalized to unity, $\int \rho(r) dr = 1$.
Jet fragmentation function $D(z)$ provides the information about the momentum spectrum of the jet fragments. It is defined as follows:
\begin{eqnarray}
D(z) = \frac{1}{N_{\rm jet}} \frac{dN_h}{dz} \,,
\end{eqnarray}
where $z =\mathbf{p}_T^{h}\cdot \mathbf{p}_T^{\rm jet} / |\mathbf{p}_T^{\rm jet}|^2$ is the momentum fraction of the jet fragments.

\begin{figure}
\begin{center}
\includegraphics*[width=12.0cm]{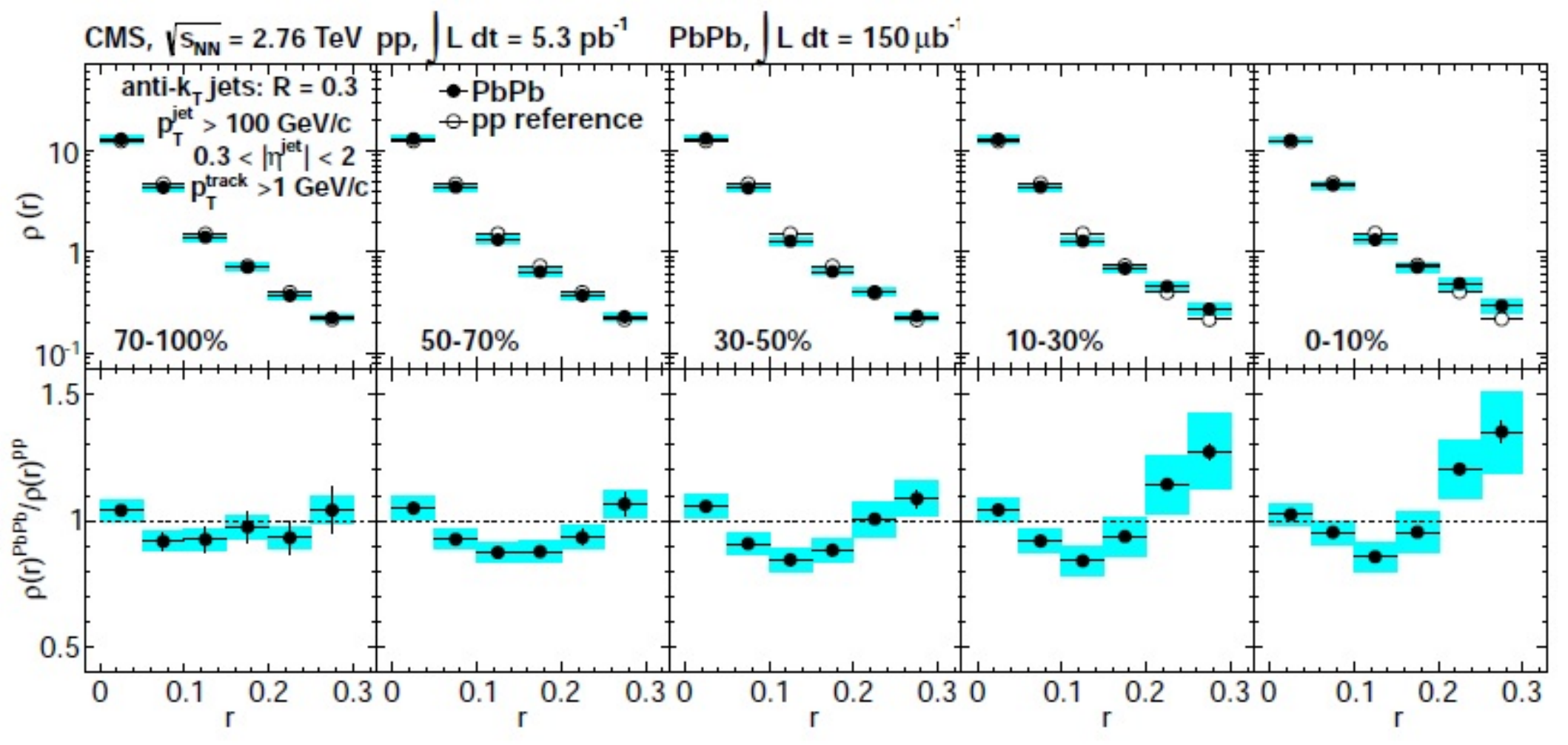}
\caption{(Color online) Top: The jet shape functions $\rho(r)$ in Pb-Pb collisions at $2.76$~ATeV for jets with $p_T^{\rm jet} > 100$~GeV/c in five centrality bins.
Bottom: The nuclear modification factors for jet shape functions.
The figures are taken from CMS measurements \cite{Chatrchyan:2012gw}.
}
\label{fig_CMS_jet_shape}
\end{center}
\end{figure}

Fig. \ref{fig_CMS_jet_shape} shows the nuclear modification of the differential jet shape function $\rho(r)$ in Pb-Pb collisions at the LHC measured by CMS Collaborations \cite{Chatrchyan:2012gw}.
The jet shape functions in peripheral Pb-Pb collisions are similar to those in p-p collisions while in more central Pb-Pb collisions we observe an excess at large radius $r > 0.2$ and a depletion at intermediate radii ($0.1 < r < 0.2$).
This indicates the broadening of the full jets after they pass through the medium.
At very small radii, little change is observed for jet shape function, suggesting that the energy distribution in the inner hard core of the jet is not affected by the jet-medium interaction.
These observations are consistent with the previous CMS finding that the lost energy from the jets is found at large distances from the jet axis \cite{Chatrchyan:2011sx}.

\begin{figure}
\begin{center}
\includegraphics*[width=12.0cm]{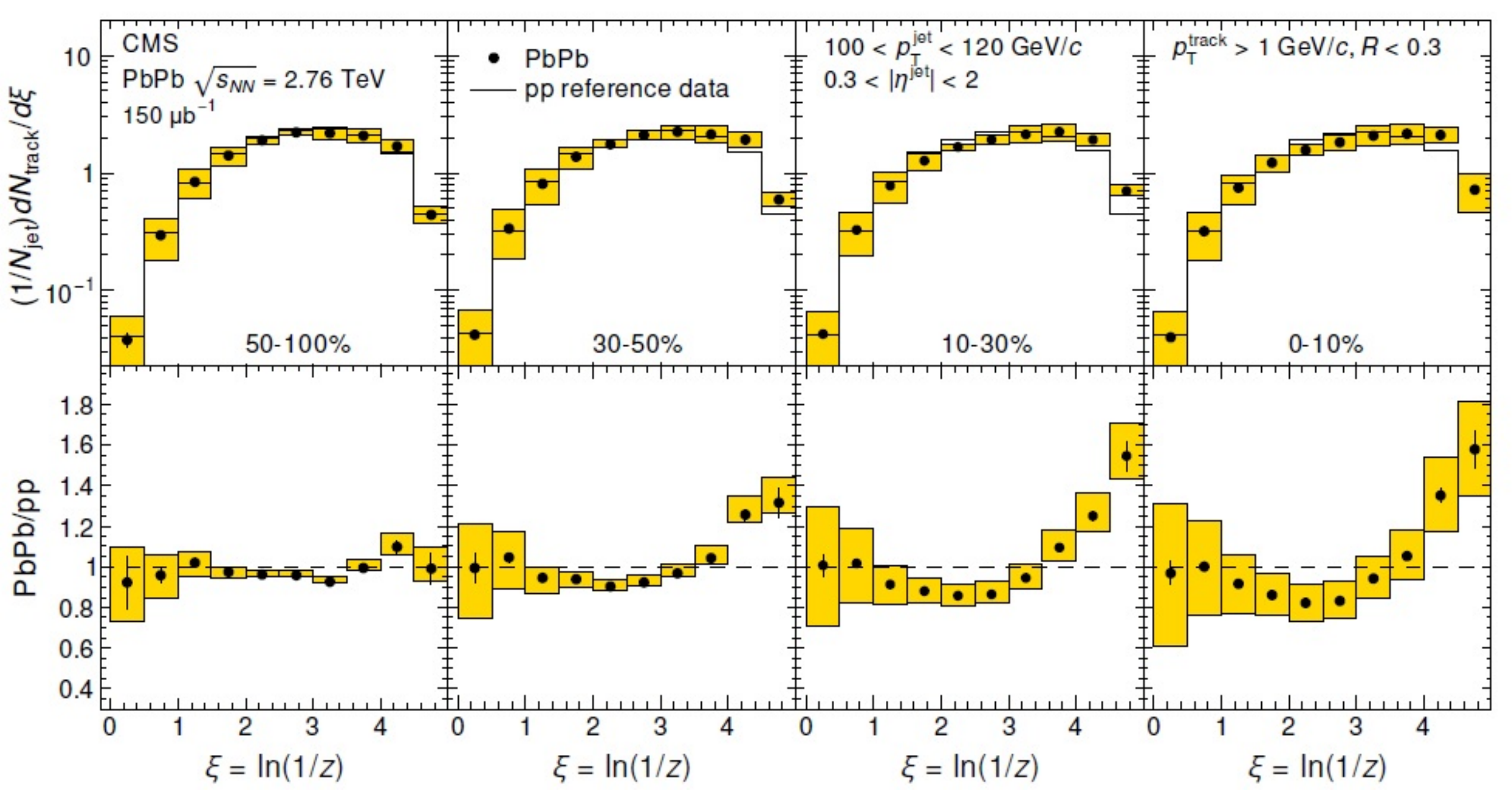}
\caption{(Color online) Top: The jet fragmentation functions $D(z)$ in Pb-Pb collisions at $2.76$~ATeV for jets with $100 < p_T < 120$~GeV/c in four Pb-Pb centrality bins.
Bottom: The nuclear modification factors for jet fragmentation functions.
The figures are taken from CMS measurements \cite{Chatrchyan:2013kwa}.
}
\label{fig_CMS_jet_ff}
\end{center}
\end{figure}

Fig. \ref{fig_CMS_jet_ff} shows the jet fragmentation functions p-p and Pb-Pb collisions at the LHC measured by CMS Collaboration \cite{Chatrchyan:2013kwa}.
Compared to the earlier CMS measurement \cite{Chatrchyan:2012gw}, the charged particles with lower values of $p_T$ have now been included in the full jet reconstruction.
We observe a clear nuclear modification of the jet fragmentation function in Pb-Pb collisions, which grows with increasing collision centrality.
In peripheral collisions (50-100\% centrality bin), the ratio of PbPb/pp is almost flat at unity.
For the most central collisions, we observe a significant excess at high $\xi=\ln(1/z)$ (low $z$) and a depletion at intermediate $\xi$.
These results indicate that the spectrum of the particles inside the full jets receives an enhancement in the soft regime, compared to p-p collisions.
ATLAS have also measured the nuclear modification of jet fragmentation function and obtained similar results: an enhanced yield of low and large $z$ fragments together with a suppressed yield of intermediate $z$ fragments.

The above observations are qualitatively consistent with jet energy loss calculations \cite{Vitev:2008rz, Zapp:2012ak, Majumder:2013re, Perez-Ramos:2014mna}.
There exist several possible sources that could affect the nuclear modification of jet substructure and contribute to the redistribution of jet energy inside the cone.
The broadening of the parton shower may be contributed from both elastic collisions with medium constituents and the medium-induced radiations.
The induced radiation may lead to the softening of the momentum spectrum of the jet fragments as well.
Particles produced from the medium response to jet transport may contribute to the redistribution of the energy inside the jets \cite{Wang:2013cia}.
Different hadronization mechanisms may lead to different jet fragmentation profiles \cite{Ma:2013gga}.
Further detailed studies on the nuclear modifications in different $r$ of jet shape distribution and different $z$ of jet fragmentation function should be able to put tighter constraint on the modeling of jet-medium interaction.

\subsection{Medium response to jet transport}

Jets may lose energy via a combination of elastic collisions and inelastic radiative processes when propagating through a QGP medium.
Some of the lost energy from the hard jets is deposited into the medium, and may contribute to many jet-related observables.
Various studies have shown that the propagating hard jets may excite the medium and produce Mach cone structure \cite{CasalderreySolana:2004qm, Ruppert:2005uz}.
If such structure is observed in relativistic nuclear experiments, it will provide a direct probe to the speed of sound of the produced nuclear matter.
However in heavy-ion collisions, the Mach cone pattern may be distorted by the large collective flow developed during the hydrodynamic expansion of the bulk matter \cite{Renk:2005si, Ma:2010dv, Bouras:2014rea, Floerchinger:2014yqa}.
The cone structure is also sensitive to various transport properties such as the shear viscosity to entropy density ratio $\eta/s$  of the QGP medium \cite{Neufeld:2008dx, Bouras:2014rea}.

One approach for studying the medium response is the use of full Boltzmann transport codes in which the jet propagation and the medium response are simulated at the same time within a single Monte-Carlo package \cite{Ma:2010dv, Bouras:2014rea}.
In this method, the bulk matter is modeled by a collection of quasi-classical partons, and the jet-medium interaction is treated the same way as the interaction among the medium constituents.
Another approach is to solve the following hydrodynamic equations,
\begin{eqnarray}
\partial_{\mu} T^{\mu \nu}(x) = J^{\nu}(x) = \left(\frac{dE}{dt}, \frac{d\mathbf{p}_\perp}{dt}, \frac{d {p}_\parallel}{dt} \right) \,.
\end{eqnarray}
Here $J^{\nu}(x)$ represents the rate of the energy and momentum deposited by the propagating hard jets.
The source term $J^{\mu}(x)$ at a given space-time location may be obtained from jet energy loss calculations, e.g., the energy deposited into the medium is equal to the collisional energy loss from the propagating jet.
For a single parton in a high temperature QGP medium, the collisional energy loss (deposition) rate in the leading logarithmic approximation may be obtained as follows:
\begin{eqnarray}
\left.\frac{dE}{dt}\right|_{\rm dep} = \left.\frac{dE}{dt}\right|_{\rm coll} = \frac{1}{4} C_s \alpha_s^2 m_D^2 \left(\frac{4ET}{m_D^2}\right) \,,
\end{eqnarray}
where $m_D^2 = 4 \pi \alpha_s (1 + N_f/6) T^2 $ is the Debye screening mass squared.

\begin{figure}
\begin{center}
\includegraphics*[width=6.1cm]{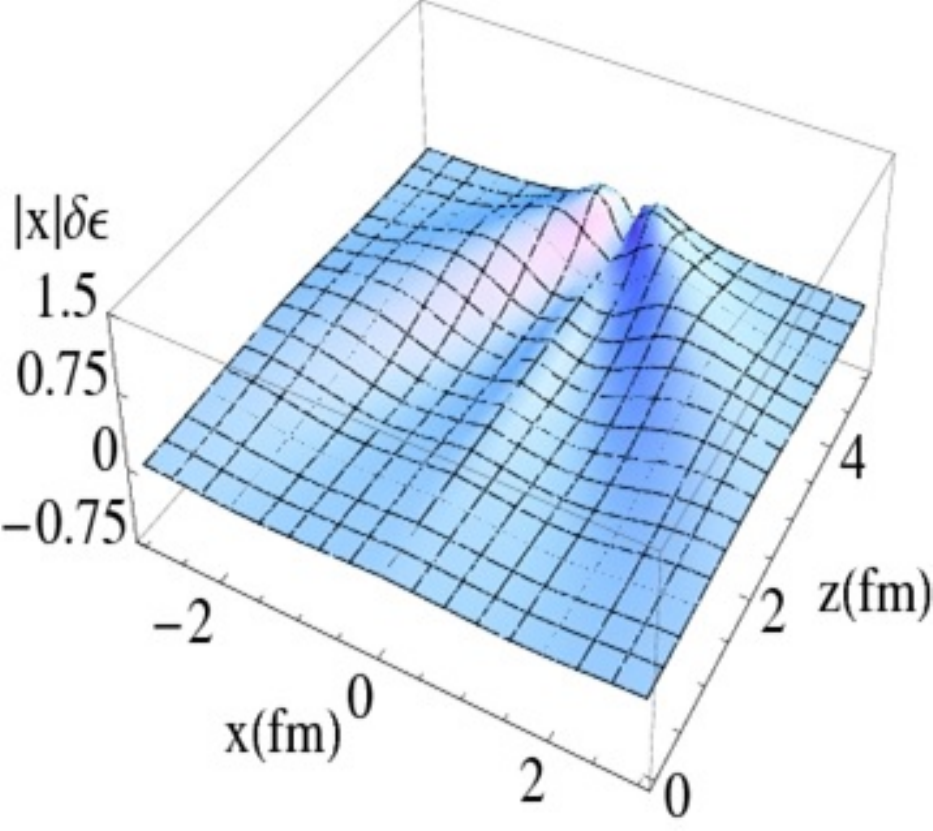}
\includegraphics*[width=6.3cm]{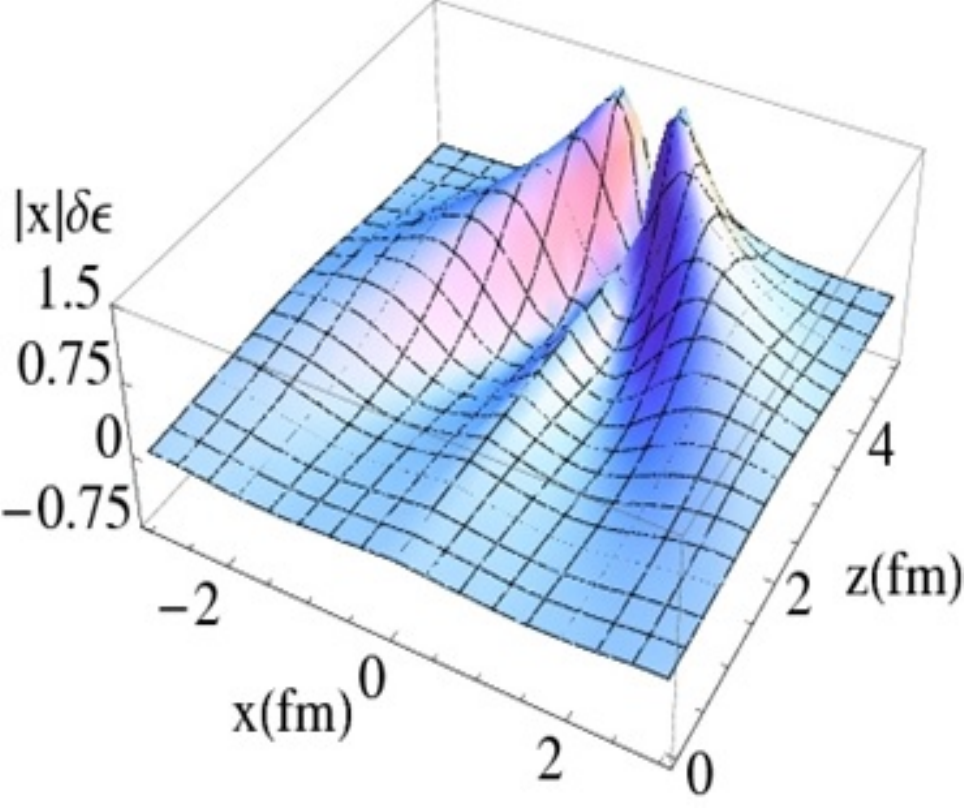}
\caption{(Color online) The medium response to the energy deposited by a primary hard quark (left panel) or by a quark-initiated parton shower (right panel) (from Ref. \cite{Qin:2009uh}).
}
\label{fig_Qin_mach_cone}
\end{center}
\end{figure}

\begin{figure}
\begin{center}
\includegraphics*[width=12.0cm]{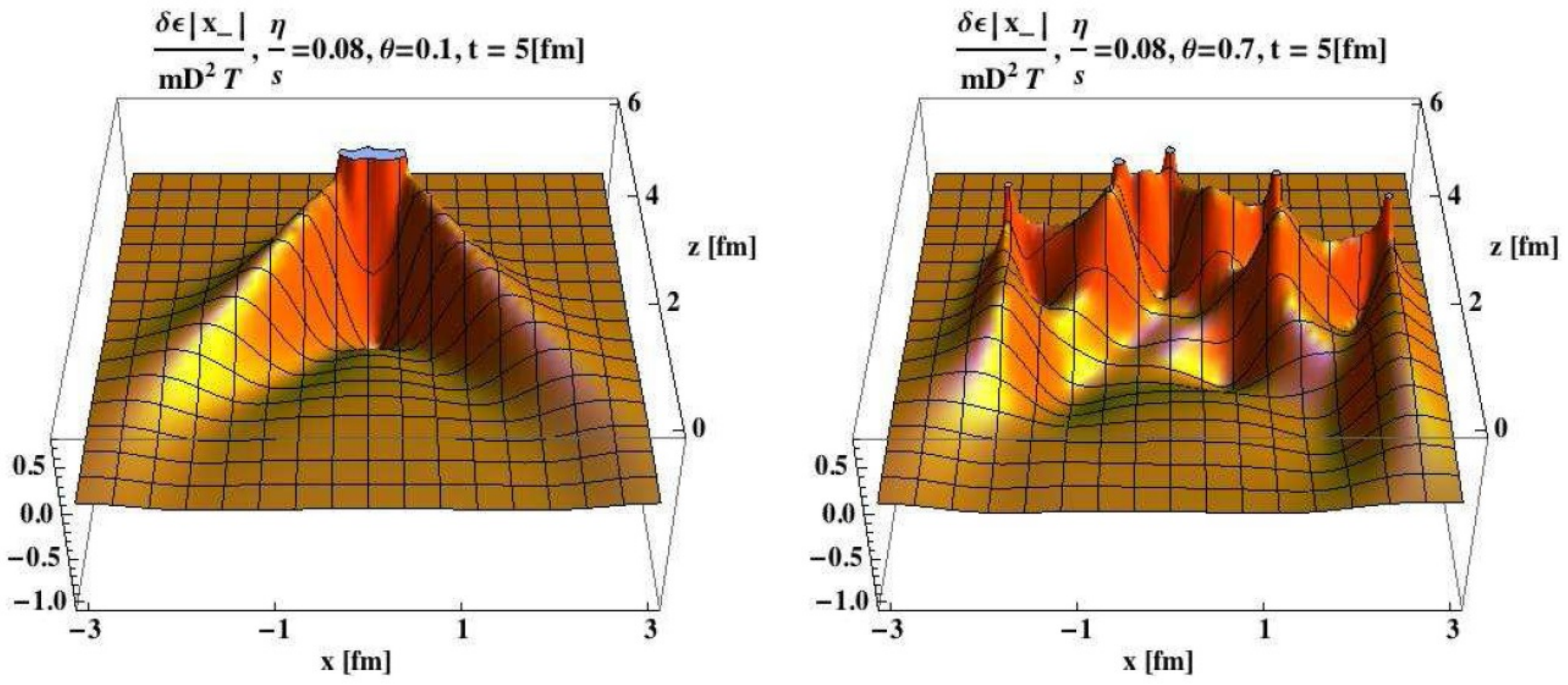}
\caption{(Color online) The medium response generated by a parton shower originating from a primary hard quark with the average gluon emission angle being $\theta = 0.1$ (left panel) and $\theta = 0.7$ (right panel), (from \cite{Neufeld:2011yh}).
}
\label{fig_Neufeld_mach_cone}
\end{center}
\end{figure}

It should be noted that jets are composed of collimated showers of partons.
Therefore, not only the leading parton, but also the radiative emissions can deposit energy and momentum into the medium via scattering with the medium constituents.
In Ref. \cite{Qin:2009uh,Neufeld:2009ep}, it is found that the length dependence of energy deposition rate by a jet shower is significantly enhanced compared to that by a single parton.
The medium responses to the energy deposited by a single parton (left panel) and by a parton shower (right panel) are compared in Fig. \ref{fig_Qin_mach_cone}.
The Mach cone structure is seen for both energy deposition cases, but the conical pattern is strongly enhanced for a parton shower.
Another important factor is the spatial distributions of the parton shower and the energy and momentum deposition profiles.
Fig. \ref{fig_Neufeld_mach_cone} compares the medium responses to a parton shower with small emission angle $\theta = 0.1$ (left panel) and with large emission angle $\theta = 0.7$ (right panel) \cite{Neufeld:2011yh}.
One can see that for nearly collinear emissions, a nice Mach cone structure can be produced.
In contrast, a well-defined Mach cone is hardly seen for the case of larger angle emissions; the medium response is now more like a superposition of several energy density perturbations.

In the above two studies (Fig. \ref{fig_Qin_mach_cone}, \ref{fig_Neufeld_mach_cone}), the medium is taken to be static, with a constant temperature (density).
In realistic event-by-event simulations which include both initial state and parton energy loss fluctuations, the energy and momentum deposition profiles may vary strongly from one event to another \cite{Renk:2013pua}.
Since the realistic medium probed by the propagating jet is dynamically evolving, expanding and cooling, the energy deposition rate will first increase and then decrease as a function of evolution time.
We also note that in many model calculations, a cutoff energy is often used to determine which part of radiation phase space ``thermalize" into the medium.
The change of such cutoff (separation) scale may lead to quite different energy loss and deposition profiles for the propagating jet \cite{Floerchinger:2014yqa,Renk:2013pua}.

\section{Summary}

One of the main goals of high energy nuclear collisions is to create the hot and dense nuclear matter with (energy) densities well above the normal nuclear matter and study its various novel properties.
In this report, we have provided some basic information and recent progresses on the study of quark-gluon plasma using relativistic heavy-ion collisions at RHIC and the LHC.
Our focus was given to the anisotropic collective flow and jet quenching phenomena, which are two most important evidences for the formation of QGP in relativistic nuclear collisions.
Below we summarize some of the main results.

From the comparison of relativistic hydrodynamics calculations with the anisotropic flow measurements, we have found that the values of shear viscosity to entropy ratio $\eta/s$ is on average larger at the LHC than that at RHIC; this indicates that the produced QGP is less strongly-coupled at the LHC.
Event-by-event fluctuation and correlation observables associated with anisotropic flow can be utilized to probe the linear as well as non-linear nature of hydrodynamic response, and can provide tight constraints on the features of the initial conditions and final state correlations of heavy-ion collisions.
The studies of longitudinal fluctuations, pre-equilibrium dynamics as well as the collective behavior in small collision systems are important for understanding the anisotropic flow phenomena in heavy-ion collisions.
Utilizing the constraint from jet quenching measurements, various phenomenological studies have shown that the average value of the scaled jet quenching parameter $\hat{q}/T^3$ is smaller at the LHC than that at RHIC; this suggests that jet-medium interaction is weaker at the LHC.
The energy loss of full jets and the nuclear modification of full jet substructures can provide us more differential and detailed information about jet-medium interaction.
The medium response to jet transport is important for more comprehensive understanding of jet-medium interaction and many jet-associated observables.

One important future task in relativistic heavy-ion collisions is to establish a general theoretical framework which not only incorporates realistic hydrodynamics models and jet energy loss/deposition calculations, but also allows us to simulate the bulk matter evolution and hard jet transport together at the same time.
Given a wealth of progresses made on the study of bulk matter and hard jets in recent years and many experimental data to come, it is hopeful that in the next few years many remaining questions can be resolved and more quantitative understanding of the transport properties of the QGP will be achieved.

\section*{Acknowledgements}
This work was supported in part by the Natural Science Foundation of China under grant No. 11375072.

%\appendix
%\section*{Appendices}

%\section*{References}

\bibliographystyle{ws-ijmpe}
%\bibliography{sample}
\bibliography{GYQ_refs}

\end{document}